\documentclass[aps,prd,showpacs,floatfix,superscriptaddress,11pt,nofootinbib]{revtex4-1}
\pdfoutput=1
\usepackage{natbib}
\usepackage[export]{adjustbox}
\usepackage{amssymb,amsmath}
\usepackage{hyperref}
\usepackage{color}
\usepackage{graphicx}
\usepackage{ulem}
\usepackage{graphics, color}
\usepackage{overpic}
\usepackage{float}
\usepackage{hyperref}
\usepackage[english]{babel}
\usepackage{hyperref}
\usepackage{amsfonts}
\usepackage[usenames,dvipsnames]{xcolor}
\usepackage{amsmath,amsfonts,amssymb,verbatim,float,mathtools}
\usepackage{physics}

\newcommand{\sigmadc}{\text{Re}(\sigma_{\rm dc})}
\newcommand{\be}{\begin{equation}}
\newcommand{\ee}{\end{equation}}
\newcommand{\bea}{\begin{eqnarray}}
\newcommand{\eea}{\end{eqnarray}}   
\newcommand{\half}{{1\over2}} 
\newcommand{\Vp}{\dot{V}}
\newcommand{\Yp}{\dot{Y}}
\newcommand{\Zp}{\dot{Z}}

\let\pa=\partial
\def\ba{\begin{array}}
	\def\ea{\end{array}}

\definecolor{Gray}{gray}{0.4}

\hypersetup{
	colorlinks   = true,
	citecolor    = blue,
	linkcolor=blue,
	urlcolor=blue
}

\def\beq{\begin{equation}}
\def\eeq{\end{equation}}

\renewcommand\Im{\operatorname{Im}}

\begin{document}
\title{Transport in a gravity dual with a varying gravitational coupling constant}
\author{Antonio M. Garc\'{\i}a-Garc\'ia}
\author{Bruno Loureiro}
\author{Aurelio Romero-Berm\'udez} 
\affiliation{Cavendish Laboratory, University of Cambridge, JJ Thomson Av., Cambridge, CB3 0HE, UK}
\begin{abstract}
We study asymptotically AdS Brans-Dicke (BD) backgrounds, where the
Ricci tensor R is coupled to a scalar in the radial dimension, as effective models of metals
with a varying coupling constant. We show that, for translational invariant
backgrounds, the regular part of the dc conductivity $\sigma_Q$ deviates from the universal result of
Einstein-Maxwell-Dilaton (EMD) models. However, the shear viscosity to entropy
ratio saturates the Kovtun-Son-Starinets (KSS) bound. Similar results apply to
more general f(R) gravity models.
In four bulk dimensions we study momentum relaxation induced by gravitational
and electromagnetic axion-dependent couplings. For sufficiently strong momentum
dissipation induced by the former, a recently proposed bound on the dc
conductivity $\sigma$ is violated for any finite electromagnetic axion coupling.
Interestingly, in more than four bulk dimensions, the dc conductivity for strong
momentum relaxation decreases with temperature in the low temperature
limit. In line with other gravity backgrounds with momentum relaxation, the
shear viscosity to entropy ratio is always lower than the KSS bound.
The numerical computation of the optical conductivity reveals a linear growth
with the frequency in the limit of low temperature, 
low frequency and large momentum relaxation. 
We have also shown that the module and argument of the
optical conductivity for intermediate frequencies are not consistent with
cuprates experimental results, even assuming several channel of momentum
relaxation.
	
\end{abstract}

\maketitle
\section{Introduction}
Einstein general relativity assumes that gravity is mediated by a tensor two
particle. Despite its immense conceptual and phenomenological success,
generalizations \cite{Goenner2012} of general relativity, where gravity is also
mediated by a scalar or a vector, have been intensively studied mostly for its
potential interest in cosmology but also simply as toy models of new ideas in
gravity. One of the most influential, though not the earliest
\cite{Goenner2012}, is the so called Brans-Dicke gravity (BD) \cite{brans1961}
that aimed to reconcile Mach's principle with general relativity. Gravity is
also mediated by a scalar coupled linearly to the Ricci tensor. The action
also has a kinetic term for the scalar so BD has two coupling constants.
Physically this scalar can be understood as a gravitation constant $G$ that
varies in time and space. General relativity is usually preferred as it predicts
the same physics with less free parameters. Interestingly, after a conformal
transformation, BD gravity maps into Einstein gravity with a dilaton field. As a
result of this mapping, explicit analytical solutions of the BD gravity
equations of motion are known not only for Einstein gravity but also for
asymptotically dS and AdS spaces even if the theory also contains massless
photons modelled by the Maxwell tensor. For a certain region of parameters it is
also possible \cite{Kang1996} to map onto BD more general f(R) models where the
action is not linear in the Ricci tensor R.

In light of this rich phenomenology, we study BD backgrounds with AdS asymptotic
as effective duals of strongly coupled metals. Previous holographic studies
\cite{sheykhi2009,sheykhi2010} involving BD backgrounds were restricted to
thermodynamic properties only. By contrast here we focus on transport
observables such as the optical, dc conductivity and shear viscosity in
asymptotically AdS Brans-Dicke backgrounds. Our motivation is to explore the
impact of the BD scalar running in the radial dimension, that acts as an
effective gravitational constant, on the transport properties of holographic
metals \cite{charmousis2010,kiritsis2015}.

More specifically we address whether the universality of the shear viscosity
\cite{policastro2001} and the regular part of the dc conductivity 
\cite{Chakrabarti2011,jain2010,Davison2015,garcia2015a}, reported in
translational invariant  Einstein-Maxwell-Dilaton (EMD)
\cite{charmousis2010,kiritsis2015} backgrounds with massless photons and no
dilaton coupling to the Maxwell tensor also holds in BD background. We have
found that, while the universal shear viscosity ratio also holds in BD
background, the finite part of the dc conductivity deviates from the universal result of EMD
theories.

We also investigate momentum relaxation by gravitational axions, namely, axions
coupled to the Ricci tensor, a simplified form of BD backgrounds where the
scalar has no dynamics.
Axions \cite{Andrade2014} together with massive gravitons, or simply a random
chemical potential
\cite{Adams2011,Arean2014,hartnoll2014,okeeffe2015,garcia2015,Araujo2016},
break translational invariance which modifies substantially the conductivity and other transport properties. For weak momentum relaxation the conductivity is to a good
extent described by Drude physics. For low temperatures or frequencies the
conductivity is large, the so called Drude peak, and decreases monotonously. It
was observed, in all models studied, that no matter the strength of the momentum
relaxation the conductivity of Einstein-Maxwell holographic metals was always
above a certain bound which precludes a metal-insulator
transition.
In part based on this numerical evidence, it was conjectured
\cite{Grozdanov2015} the existence of a lower bound in the conductivity of more
complicated holographic models models. However, in two recent papers
\cite{Baggioli2016,Gouteraux2016} violations of this bound have been reported in
models where the axion is coupled to the Maxwell tensor, effectively screening
charge.
Here we show that gravitational axions, that do not screen charge, also lead to
violations of the bound in the limit of strong, although still parametrically
small with respect to the rank of the gauge group, axion gravitational coupling.
For three space dimensions the dc conductivity decreases with $T$ for low
temperatures even without any other source of momentum relaxation.

We also study the optical conductivity in BD backgrounds.
The optical conductivity in EMD models with translational invariance in the
limit of small frequencies and temperatures is controlled by the infrared (IR)
geometry that for Reissner-Nordstrom background is AdS$_2$ leading to $\sigma
\sim \omega^2$.
The effect of momentum relaxation in the optical conductivity of EMD theories
was investigated in \cite{kiritsis2015} but it is not yet fully understood
whether, for low-frequencies and strong momentum relaxation, the conductivity
scales as a power-law faster than linear as in Mott insulators and many-body
localised states \cite{demler2015}. 

By contrast, in a model in which momentum relaxation occurs by a oscillatory chemical potential, it was claimed \cite{horowitz2013} the modulus of
the optical conductivity for intermediate frequencies decays as a power-law with
an exponent equal to that observed in most cuprates. Here we find that, even
assuming several channels of momentum dissipation, we cannot reproduce the
modules and argument observed in cuprates. However, we have found that, for
strong momentum dissipation and close to zero temperature, the optical
conductivity increases linearly, not quadratically with the frequency, for both
gravitational and electromagnetic axions. Finally, we have computed the ratio of the
shear viscosity and the entropy density in BD holography with momentum
relaxation. We have observed that, unlike the translational invariant case, the ratio is
temperature dependent. It decreases as the strength of momentum relaxation
increases and it is always below the KSS bound. It can be made arbitrarily small
for a finite amount of momentum relaxation.

The organization of the paper is as follows: in section two we compute
analytically the regular part of the dc conductivity and show that the shear viscosity to entropy
ratio in translational invariant BD backgrounds and other generalized theories
of gravity is given by the KSS bound. In section three we study the dc
conductivity in BD like backgrounds with momentum relaxation induced by coupling
the axion and the Ricci tensor in two boundary space dimensions.
In section four we address momentum relaxation by gravitational axions in higher space
dimensions.
In section five we study the optical conductivity in BD backgrounds. We also
compute the module and argument of the complex conductivity in order to compare
with results in cuprates.
In section six we compute the shear viscosity to entropy density ratio including
different sources of momentum relaxation. We end up with a list of interesting
problems for further research and a short summary of the main results.

Next we introduce the BD action, the equations of motion and its analytical
solution.

\section{dc Conductivity in translationally invariant BD holography}
We start our analysis by introducing the BD action and the equations of motion
(EOM). We then compute the conductivity for a general background ansatz and show
it is expressed in terms of thermodynamic quantities and the value
of the scalar at the horizon. This is different from EMD models with no coupling
between the dilaton and the Maxwell field where it only depends on thermodynamic
quantities.

We then find that a calculation in the Einstein frame, resulting from a
conformal transformation, leads to the same result. Finally, we discuss other
modified gravity models that fall within the BD universality class.

\subsection{Brans-Dicke Action and equations of motion}
The Brans-Dicke-Maxwell action in a $d+1$-dimensional manifold is given by
\begin{align}
\label{action}
S=\int_{\mathcal{M}}\dd[d+1]x\sqrt{-g}\left[\phi R -\frac{\xi}{\phi}(\nabla\phi)^2-V(\phi)-\frac{Y}{4} F^2 \right].
\end{align}
Note that we are working in units where $2\kappa^2 =16\pi G_N = 1$ and we include a non-trivial coupling $Y(\phi)$ between the Brans-Dicke scalar $\phi$ and the Maxwell term, as well as a (for now arbitrary) scalar potential $V(\phi)$. In this model gravity is not only mediated by the massless symmetric rank two tensor $g$ but also by the real scalar field which has its own dynamics and a kinetic term parametrized by $\xi\geq 0$\footnote{The standard notation in the literature for the Brans-Dicke coupling is $\omega$. We refrain from this notation to avoid confusion with the frequency $\omega$ in the optical conductivity $\sigma(\omega)$.}. Intuitively the non-minimal coupling $\phi R$ can be interpreted as the running of Newton's constant ``$ G(x) \equiv G_N/\phi(x)$'' \cite{Kang1996}.

Variation of this action gives the following EOM's:
\begin{align*}
\phi\left(R_{ab}-\frac{1}{2}R g_{ab}\right) =& \frac{\xi}{\phi}\left(\nabla_a\phi\nabla_b\phi-\frac{1}{2}(\nabla\phi)^2~g_{ab} \right)-\frac{1}{2}V(\phi)+(\nabla_a\nabla_b\phi-\square \phi~g_{ab})\\
&-\frac{Y}{2}\left(F_{ac}F^{c}_{\phantom{c}b}+\frac{1}{4}F^2~g_{ab}\right), \\
\partial_a\left(\sqrt{-g}Y(\phi)F^{ab}\right)&=0,\\
\square\phi &= \frac{1}{2(d-1)\xi+2d}\Big((d-1)\phi V'(\phi)-(d+1)V(\phi)-\frac{(d-3)}{4}F^2-\frac{Y'(\phi)}{4}\phi F^{2} \Big).
\end{align*}
First, note that there is an extra term $(\nabla_a\nabla_b\phi-\square
\phi~g_{ab})$ for the scalar in the Einstein's equations. This term comes from
Palatini identity $\delta_g R = R_{ab}\delta g^{ab}-\nabla_c
\left(\delta\Gamma^{c}_{ab}-g^{ac}\delta\Gamma^{b}_{cb}\right)$ which in
Einstein gravity can be integrated to give a boundary term. Evaluating this term
in normal coordinates and using Stoke's theorem yields the extra term previously
mentioned. Second, terms on the left hand side of the scalar equation come from
the Ricci scalar factor that is solved by taking the trace of Einstein's
equations.

An important observation is that for $Y=1$ and $d=3$ the Maxwell term in the
scalar equation vanishes. This is a consequence of the fact that in $d=3$ the
electromagnetic energy-momentum tensor is conformal and therefore traceless, and
do not source the Ricci scalar. This invariance will play later a crucial role in our
analysis.

\subsection{Regular part of the dc Conductivity for a General Ansatz}\label{sec:dc_BD}
We now present the computation of the conductivity at zero frequency in a generic static and spherically symmetric
AdS$_{d+1}$ black brane. As usual in translationally invariant theories, for vanishing frequency $\sigma(\omega\to 0)
\to \sigma_Q+K \delta(\omega)$, where we use the standard notation \cite{Davison2015}: $K$ for the Drude weight and 
$\sigma_Q$ for the regular part of the dc conductivity. 
In this section we are only interested in the latter.
We derive a general expression for $\sigma_Q$ that highlights the universality of our results. An explicit
solution is worked out in appendix \ref{app1}.

\subsubsection{Background and conserved charges}
Consider the following static and spherically symmetric ansatz for the field equations,
\begin{align}\label{eq:ansatz}
\dd s^2 &=-A(r)\dd t^2+B(r)\dd r^2+C(r)\delta_{ij}\dd x^i \dd x^j,\\
A &= a_{t}(r)\dd t,\\
\phi &= \phi(r).
\end{align}
We assume this chart is globally defined and describes an asymptotically AdS$_{d+1}$ black hole. More precisely, we require that $A(r)=B(r)^{-1} = C(r) = r^2$ as $r\to\infty$ (assymptotic boundary) and that $A(r)\sim B(r)^{-1} \sim 4\pi T (r-r_0)$ for $r_0>0$. In what concerns the fields, we need to require that $Y(\infty)=1$, $\phi(\infty)\neq 0$ and $V(\phi) \sim 2\Lambda \phi$ close to the boundary and that they are regular at the horizon $r=r_0$. Moreover we impose $a_t(r_0)=0$.

For this ansatz we use the existence of two radially conserved charges in order to simplify the computation of the conductivity. The simplest one is the charge density, that can be obtained by looking at the $(t)$ component of Maxwell's equations,
\begin{align*}
\partial_a\left(\sqrt{-g}Y(\phi)F^{at} \right) =
\partial_r\left(\sqrt{-g}Y(\phi)g^{rr}g^{tt}a_{t}' \right) =0\ .
\end{align*}
Therefore the charge density 
\begin{align}
\label{charge1}
\rho = \frac{Y C^{\frac{d-1}{2}}}{\sqrt{AB}}a_{t}'
\end{align}
is radially conserved. The second conserved quantity is related to the geometry. Consider the $(tt)$ and the $(xx)$ components of the Brans-Dicke equations with one raised index
\begin{align*}
\phi\left(R^{t}_{\phantom{t}t} - \frac{1}{2}R ~g^{t}_{\phantom{t}t}\right)
=-\frac{Y}{2}F_{tr}F^{tr}+\left(-\frac{Y}{8}
F^2+\frac{\xi}{2\phi}(\nabla\phi)^2+\frac{1}{2}V-\square\phi\right)g^{t}_{\phantom{t}t}+\nabla^{t}\nabla_{t}\phi\ ,\\
\phi\left(R^{x}_{\phantom{x}x} - \frac{1}{2}R ~g^{x}_{\phantom{x}x}\right) =
\left(-\frac{Y}{8}
F^2+\frac{\xi}{2\phi}(\nabla\phi)^2+\frac{1}{2}V-\square\phi\right)g^{x}_{\phantom{x}x}+\nabla^{x}\nabla_{x}\phi\ .\\
\end{align*}
Note that the assumption  $\partial_t \phi = \partial_x\phi =0$ is crucial in this analysis. For our ansatz, $g^{t}_{\phantom{t}t}=g^{x}_{\phantom{x}x}=1$ and therefore we subtract the above expression to give
\begin{align}
\label{conserved1}
\phi\left(R^{t}_{\phantom{t}t} -
R^{x}_{\phantom{x}x}\right)=-\frac{Y}{2}F_{rt}F^{rt}+\nabla^{t}\nabla_{t}\phi -
\nabla^x\nabla_x\phi\ .
\end{align}
In order to write this as a total derivative, we will make use of the following identities
\begin{align*}
\sqrt{-g}R^{t}_{\phantom{t}t}=-\frac{1}{2}\partial_r\left(\frac{1}{\sqrt{AB}}C^{\frac{d-1}{2}}A'\right),\\
\sqrt{-g}R^{x}_{\phantom{x}x}=-\frac{1}{2}\partial_r\left(\sqrt{\frac{A}{B}}C^{\frac{d-3}{2}}C'\right).
\end{align*}
Which is subtracted to give
\begin{align*}
\sqrt{-g}(R^{t}_{\phantom{t}t}-R^{x}_{\phantom{x}x})=-\frac{1}{2}\partial_r\left(\frac{1}{\sqrt{AB}}C^{\frac{d+1}{2}}\left(\frac{A}{C}\right)'\right).
\end{align*}
Regarding the right hand side, we have
\begin{align*}
\nabla^t\nabla_t \phi - \nabla^x\nabla_x \phi = g^{tt}\nabla_t\nabla_t \phi - g^{xx} \nabla_x\nabla_x \phi = -\left(g^{tt}\Gamma^{r}_{tt}-g^{xx}\Gamma^{r}_{xx}\right)\phi' = \frac{1}{2B}\left(\frac{A'}{A}-\frac{C'}{C}\right)\phi',
\end{align*}
and therefore
\begin{align*}
\sqrt{-g}\left(\nabla^t\nabla_t \phi - \nabla^x\nabla_x \phi \right)= \frac{C^{\frac{d+1}{2}}}{2\sqrt{AB}}\left(\frac{A}{C}\right)'\phi'.
\end{align*}
Taking into account that $\sqrt{-g}YF^{rt}=\rho$, we finally rewrite eq. \eqref{conserved1} as
\begin{align}
\label{conserved2}
\phi~\partial_r\left(\frac{1}{\sqrt{AB}}C^{\frac{d+1}{2}}\left(\frac{A}{C}\right)'\right)=\rho a_t' - \frac{C^{\frac{d+1}{2}}}{\sqrt{AB}}\left(\frac{A}{C}\right)'\phi'.
\end{align}
which is expressed as a total derivative,
\begin{align}
\label{conserved3}
\partial_r\left(\frac{\phi}{\sqrt{AB}}C^{\frac{d+1}{2}}\left(\frac{A}{C}\right)'-\rho
a_t\right)=0\ .
\end{align}
In previous works, this conserved charge was related to thermodynamical quantities \cite{jain2010, Chakrabarti2011, Davison2015}. Indeed, by integrating it, evaluating it at the horizon and using the boundary condition $a_t(r_0)=0$ we get
\begin{align*}
\left. \frac{\phi}{\sqrt{AB}}C^{\frac{d+1}{2}}\left(\frac{A}{C}\right)'\right|_{r=r_0} = \frac{\phi(r_0)}{\sqrt{AB}}C(r_0)^{\frac{d-1}{2}}A'(r_0) = sT
\end{align*}
\noindent which is exactly the same results obtained previously. Note that we use the fact that the entropy in BD theory satisfies
\begin{align*}
S = \frac{1}{4}\int_{r=r_0}\dd[d]x~\phi\sqrt{-g}
\end{align*}
 instead of the standard area law. 
This is another simple manifestation that the strength of gravity in BD backgrounds is not constant, as in Einstein gravity, but governed by the scalar ``$G/\phi$''\footnote{Note that in standard units the area law is $S=\frac{1}{4G}\int_{r=r_0}\dd[d]x \sqrt{-g}$} \cite{Kang1996}. Surprisingly, the scalar field fits nicely to produce the same thermodynamic result.

For a general $r$, we then have
\begin{align}
\label{charge2}
sT = \frac{\phi}{\sqrt{AB}}C^{\frac{d+1}{2}}\left(\frac{A}{C}\right)'-\rho a_t\
.
\end{align}
Note that in particular $a_t(\infty)=\mu$.  Therefore using the Smarr relation $sT+\mu\rho = \epsilon + P$ and the above expression evaluated at the boundary $r\to\infty$ we get
\begin{align*}
\epsilon+P=\left. \frac{\phi}{\sqrt{AB}}C^{\frac{d+1}{2}}\left(\frac{A}{C}\right)' \right|_{r=\infty}.
\end{align*}

\subsubsection{Fluctuations}
In order to compute conductivities, we need to study fluctuations around the background solutions. It is sufficient to consider the following set of consistent fluctuations
\begin{align*}
\dd s^2\to \dd s^2 + 2h_{tx}(t,r)\dd t \dd x\ , && A\to A+a_{x}(t,r)\dd x\ .
\end{align*}
The EOM for $a_x$ is obtained by linearizing the $(x)$ component of Maxwell's equations,
\begin{align*}
\partial_a\left(\sqrt{-g}Yg^{ac}g^{xd}F_{cd}\right) &= \partial_{r}\left(\sqrt{-g}Yg^{rr}g^{xx}\partial_r a_x\right)-\partial_{r}\left(\sqrt{-g}Yg^{rr}g^{xt}a_t'\right)+\partial_{t}\left(\sqrt{-g}Yg^{tt}g^{xx}\partial_t a_x\right)\\
&=\partial_r\left(\sqrt{\frac{A}{B}}C^{\frac{d-3}{2}}Y\partial_r
a_x\right)-\sqrt{\frac{B}{A}}C^{\frac{d-3}{2}}Y\partial_t^2a_x+\rho\partial_r\left(g^{xx}
h_{tx}\right)=0\ .
\end{align*}
To eliminate the $h_{tx}$ term in the above expression we look at the constraint equation given by the linearized $(rx)$ Brans-Dicke equation,
\begin{align*}
\partial_r\left(g^{xx}h_{tx}\right)=-\frac{g^{xx}}{\phi}Y a_t'a_x\ .
\end{align*}
Inserting this in the above expression and solving for $a_t'$ in function of the charge density we get
\begin{align*}
\frac{\phi~
C^{\frac{d+1}{2}}}{\sqrt{AB}}\partial_r\left(\sqrt{\frac{A}{B}}C^{\frac{d-3}{2}}Y\partial_r
a_x\right)-\sqrt{\frac{B}{A}}C^{\frac{d-3}{2}}Y\partial_t^2a_x-\rho^2 a_x=0\ .
\end{align*}
We now use the conserved charge \eqref{conserved3} to rewrite the above equation as,
\begin{align*}
\partial_r\left(\frac{C^{d-1}Y\phi}{B}\left(\frac{A}{C}\right)' \partial_r
a_x-\frac{A}{C}\rho^2
a_x\right)-\sqrt{\frac{B}{A}}C^{\frac{d-3}{2}}Y\partial_t^2a_x=0\ .
\end{align*}
As long as we are ony interested in the regular part of the dc conductivity, we can set $\partial_t^2 a_x=0$. The resulting equation is easily integrated to give,
\begin{align*}
\frac{C^{d-1}Y\phi}{B}\left(\frac{A}{C}\right)' \partial_r a_x-\frac{A}{C}\rho^2
a_x = \text{constant}\ .
\end{align*}
Black brane boundary conditions set $A(r_0) \sim 1/B(r_0)=0$. Regularity of the fields at the horizon fixes the constant above to zero. Moreover,without loss of generality we set $\lim\limits_{r\to\infty}\phi(r) \equiv 1$ at the boundary. With this information the equation for the fluctuation is easily integrated: 
\begin{align*}
a_x^{(0)}(r)=\exp\left\{-\int_{r}^{\infty} \frac{AB \rho^2}{Y\phi
(A/C)'C^{d-2}}\dd r \right\} =\exp\left\{-\int_{r}^{\infty}
\frac{Y(a_t')^2}{\phi C(A/C)'}\dd r \right\}\ ,
\end{align*}
\noindent where $a^{(0)}_x(r)$ is the independent solution of the equation that tends to one at the boundary and determines the regular part of the conductivity \footnote{The second solution can be obtained using the Wronskian. For further details see \cite{Davison2015}.}. We can use the charges \eqref{charge1} and \eqref{charge2} to perform the integral explicitly:
\begin{align*}
\int_{r}^{\infty} \frac{Y(a_t')^2}{\phi C(A/C)'}\dd r = \int_{r}^{\infty}
\frac{a_t'}{a_t + sT/\rho}\dd r = \left. \log( a_t(r)\rho +
sT)\right|_{r}^{\infty} = \log\frac{\epsilon+P}{a_t(r)\rho+sT}\ ,
\end{align*}
\noindent where we have used the Smarr relation. This implies that
\begin{align*}
a^{(0)}_x(r)=\frac{a_t(r)\rho+sT}{\epsilon+P}\ ,
\end{align*}
\noindent and in particular
\begin{align*}
a^{(0)}_x(r_0) = \frac{sT}{\epsilon+P}\ ,
\end{align*}
\noindent where we use the boundary condition $a_t(r_0)=0$. Finally, the regular part of the dc conductivity is given by
\begin{align*}
\sigma_Q=Y(r_0)C(r_0)^{\frac{d-3}{2}}(a_x^{(0)}(r_0))^2 =
Y(r_0)C(r_0)^{\frac{d-3}{2}}\left(\frac{sT}{\epsilon+P}\right)^2.
\end{align*}
This result is the same as the one obtained in EMD models \cite{jain2010, Chakrabarti2011, Davison2015}. It is interesting to observe that in these works $T^{t}_{t} = T^{x}_{x}$ is given as a necessary condition for the universality of $\sigma_Q$. Here this condition is clearly violated, but we can still write the equation for the fluctuation as a total derivative. However, taking into account the modified area law for the entropy density,
\begin{align}\label{conduc}
\sigma_Q = \frac{Y(r_0)}{\phi(r_0)^{\frac{d-3}{d-1}}}
\left(\frac{s}{4\pi}\right)^{\frac{d-3}{d-1}}\left(\frac{sT}{\epsilon+P}\right)^2,
\end{align}
\noindent we highlight the explicit dependence of $\sigma_Q$ on the BD field $\phi$. For $d>3$, we thus expect the scalar field to renormalize the universal contribution to the conductivity. To understand better the possible effects of $\phi$, we need to evaluate its behavior at the horizon, in particular, the low and high temperature scaling of $\phi(r_0)$.

In contrast with the previous discussion, those questions do not have a universal answer. It depends on the particular  solution for the background. In next section, we will see that the BD model can be formally mapped onto an EMD model by a conformal transformation. Explicit solutions for the latter have been widely studied in both the context of gravity \cite{Gibbons1988,Garfinkle1991} and of holography \cite{charmousis2010, Gouteraux2011a} (and references therein). As an illustration of this method we study in appendix \ref{app1} a particular solution, and show explicitly how it is mapped to BD.

A simple numeric fit for the solution \eqref{bdsolution} suggests two scaling regimes. For low temperature, $\phi(T)\sim a$ tends, for a fixed charged density, to a temperature independent constant $0<a<1$. This indicates that close to extremality the temperature scaling of $\sigma_Q$ coincides with EMD result, although up to a numerical prefactor $a$. For high temperatures, we find that $\phi(T) \sim T^{-\delta}$ for $\delta \geq 0$, suggesting that $\phi\to 0$ asymptotically at the horizon for sufficiently high temperatures. Here $\delta$ is a function of both the Brans-Dicke parameter $\xi$ and the dimensionality $d$, and monotonically decreases with $\xi\geq 0$ for fixed $d$. For example, for $d=4$ and $\xi=0$, we have $\delta \approx 0.4$, while $\delta\approx 0.22$ for $\xi=1$. In particular we have $\delta\to 0$ as $\xi\to\infty$ for any dimension $d>2$.

One can interpret this behavior in a heuristic way. First note that Newton's constant $G$ is related to the string coupling constant $g_s \sim G$. If we naively interpret the BD coupling ``$G/\phi$'' as a dynamical Newton's constant, the flow of $\phi$ can be interpret as a flow from weaker ($\phi\gg 1$) to stronger ($\phi \ll 1$) coupling. More specifically, for the background solution of appendix \ref{app1} $\phi\in [0,\phi(r_0)]$ with $0< \phi(r_0)\leq 1$. Therefore the running of $\phi$ from the boundary to the horizon corresponds in the dual field theory to a flow from weaker to stronger coupling.  For the purpose of the conductivity, this running has the effect of increasing $\sigma_Q$ by a factor $\phi(r_0)^{-\frac{d-3}{d-1}}$. Although tempting, one needs to be cautious with this heuristic interpretation. In the saddle point approximation, exact only in $ N \to \infty$ limit, we always have $g_s \ll 1$ and $\lambda = g_s N\gg 1$. Thus this interpretation should not be taken seriously in the limit of fixed large $N$ and $\phi(r_0)\to 0$, where the saddle point is clearly not valid.\footnote{For this reason we employ the term 'weaker' instead of `weak'.}

\subsubsection{Conformal transformations and Universality}
\label{sec:conf_transf} 
The explicit result for the regular part of the dc  conductivity (\ref{conduc}) is also expected from a well known trick broadly used in the Brans-Dicke literature that we now discuss (see for example \cite{Brans2005} and references within). 

Consider the following conformal mapping of the metric $g$,
\begin{align*}
\bar{g} = \phi^{\frac{2}{d-1}}g\ .
\end{align*}
Taking into account the transformation in the volume element and in the Ricci
scalar, the action reads
\begin{align*}
\bar{S} = \int_{\mathcal{M}}\dd[d+1]x\sqrt{-g}\left[\bar{R}-\frac{4}{d-1}(\bar{\nabla}\bar{\phi})^2-\bar{V}(\bar{\phi})-\frac{\bar{Y}(\bar{\phi})}{4}\bar{F}^2 \right],
\end{align*}
\noindent where we have defined
\begin{align}\label{eq:coup_conf_transf}
\alpha = \frac{d-3}{2\sqrt{(d-1)\xi+d}} && \bar{\phi}=\frac{d-3}{4\alpha}\log{\phi}\\
\bar{V}(\bar{\phi})=\phi^{-\frac{d+1}{d-1}}V(\phi) && \bar{Y}(\bar{\phi})=Y(\bar{\phi})e^{-\frac{4\alpha\bar{\phi}}{d-1}},
\end{align}
\noindent and all bar $\bar {\ldots}$ quantities are computed with respect to the metric $\bar{g}$. Note that for $d=3$ the Maxwell coupling is not affected by the conformal mapping. This is a consequence of the fact that electromagnetism is conformal in $d=3$. It is also useful to note that $\bar{\phi}$ is well defined for $d=3$ since $\alpha$ has a factor $d-3$ as well.

This is nothing but the well known Einstein-Maxwell-Dilaton action. This action has been widely studied in the context of string theory and effective holographic models \cite{gao2005, sheykhi2010, charmousis2010, kiritsis2015}. This map provides a useful way of constructing solutions to BD gravity, since solutions of EMD theory are well known. A particular explicit solution that illustrates this point is given in appendix \ref{app1}. More interestingly, it is  known that the regular part of the dc conductivity in such models take the (almost-)universal form
\begin{align*}
\bar{\sigma}_Q = \bar{Y}(r_0) \left(\frac{\bar{s}}{4\pi}\right)^{\frac{d-3}{d-1}}\left(\frac{\bar{s}\bar{T}}{\bar{\epsilon}+\bar{P}}\right)^2.
\end{align*}
It is not hard to check that the thermodynamic quantities $(\bar{s}, \bar{T}, \bar{\epsilon}, \bar{P})$ are invariant under the conformal mapping \footnote{This is essentially a consequence of the regularity of $\phi$ at the horizon.}. The only part of $\bar{\sigma}_Q$ that is not invariant is the non-universal charge coupling $\bar{Y}$, which transforms as $\bar{Y}(\bar{\phi}) = \phi^{-\frac{d-3}{d-1}}Y(\phi)$ and give the explicit result computed in the previous section.

This raises the interesting question of whether there are other theories of modified gravity that can be cast as an EMD theory in the Einstein frame, and if so, which those theories are. Indeed, this question has been much discussed in the gravity literature \cite{Shapiro1995, Shapiro1997, Sotiriou2010}. There has been a controversial debate on whether theories that are related by field redefinition or conformal transformation describe {\it gravitationally} \footnote{By gravitationally we mean the geodesics, conservation of the energy-momentum tensor, energy conditions, etc.} the same theory or not \cite{SOTIRIOU2008b}. A full discussion of those intrincate questions is beyond the scope of the paper. Here we limit our discussion to the fact that conformal transformations are a convenient tool to study dynamically equivalent theories.

A well known class of theories that can be mapped into BD are $f(R)$ theories of gravity, defined through the action
\begin{align*}
S = \int_{\mathcal{M}}\dd[d+1] x \sqrt{-g}f(R) + S_{\text{matter}}\ ,
\end{align*}
\noindent where $f(R)$ is a generic function of the Ricci scalar $R$ and $S_{\text{matter}}$ include any other fields coupled to the metric, but not to $R$. One can introduce an auxiliary field $\chi$ to rewrite the above equation as
\begin{align*}
S = \int_{\mathcal{M}}\dd[d+1] x \sqrt{-g}\left[f(\chi)+f'(\chi)(R-\chi)\right]
+ S_{\text{matter}}\ .
\end{align*}
Variation with respect to $\chi$ give $f''(\chi)\left(R-\chi\right)=0$, so as long as $f''(\chi)\neq 0$ this constraint imposes $\chi = R$ and we recover the initial action. Note that this is also a sufficient condition for $f(R)$ to be invertible. Now defining $\phi = f'(\chi)$ and $V(\phi) = \chi(\phi)\phi-f(\chi(\phi))$ we can write
\begin{align*}
S = \int_{\mathcal{M}}\dd[d+1] x \sqrt{-g}\left(\phi R - V(\phi)\right) +
S_{\text{matter}}\ ,
\end{align*}
\noindent which is precisely a BD action with $\xi=0$ and potential $V(\phi)$. This procedure is nothing but the Legendre transform of the action with respect to the conjugate field $\phi$. Taking $S_{\text{matter}}=-\frac{1}{4} \int Y(r)F^2$, we proceed with the computation of the regular part of the conductivity as before to get
\begin{align*}
\sigma_Q = \frac{Y(r_0)}{f'(R(r_0))^{\frac{d-3}{d-1}}} \left(\frac{s}{4\pi}\right)^{\frac{d-3}{d-1}}\left(\frac{sT}{\epsilon+P}\right)^2.
\end{align*}
This illustrates how a combination of a conformal transformation and a Legendre
transform can be used to considerably simplify calculations. Indeed, this
procedure is much more general, and can be applied to other theories such as
Palatini gravity or $f(\phi)$ couplings to the Ricci \cite{SOTIRIOU2008b}. An
interesting example of the latter is a conformal coupling $f(\phi) =
1+\frac{1}{6}\phi^2$ that appears naturally in one-loop diagrams of string theory
\cite{Shapiro1995a}.

It is tempting to apply this construction to other theories of gravity such as
Gauss-Bonnet, which in holography effectively correspond to leading $1/N$ corrections in the dual field theory. However, Gauss-Bonnet contain terms such as $R_{ab}R^{ab}$ which introduce further non-linearity and thus makes difficult the Legendre transformation \cite{Sotiriou2010}. Therefore Gauss-Bonnet gravity do not fall under BD universality.

A natural question to ask is what happens with other transport coefficients. For
example, both the shear viscosity and the entropy contain the same power of $G$,
and therefore $\eta/s$ does not depend on $G$. As a consequence, we expect $\eta/s = 1/4\pi$ in BD holography to saturate the KSS bound. This is just a particular example of
a general result that any theory related to standard gravity via a conformal
transformation indeed saturates the KSS bound \cite{Brustein2009, Brustein2009a}.
However, quantities such as the entanglement entropy should be sensitive to
$\phi$ in the expected way ($G \to G\phi^{-1}$). Indeed this was explicitly
calculated in the context of $f(R)$ theories, and agrees with our discussion
since $\phi=f'(R)$ \cite{Nojiri2011,Chatterjee2012, Pourhasan2014}.

This discussion applies only to theories with no momentum relaxation. In the rest of the paper we focus on the description of transport in BD holography with momentum dissipation. 

\section{Momentum relaxation and dc conductivity in BD holography}

We now study the effect of momentum relaxation in the transport properties of the field theory dual of BD gravity. 
We consider the linear coupling to the Ricci scalar to be a function of the gradient of axion fields that explicitely break diffeomorphism invariance in the boundary spacelike coordinates:
\begin{equation}\label{eq:action}
S=\int \dd[d+1]x \sqrt{-g}\left[Z(TrX)R-2\Lambda-V(TrX)-{Y(TrX)\over 4} F^2\right]\ ,
\end{equation}
where $TrX={1\over d-1}\sum_I \nabla_\mu X^I\nabla^\mu X^I$ and $X^I=\alpha
x^I$. We use the metric ansatz of eq. (\ref{eq:ansatz}) with
$A=g$, $B=1/g$ and $C=r^2c$:
\begin{equation}\label{eq:metric}
\dd s^2=-g(r)\dd t^2+{\dd r^2\over g(r)}+r^2c(r)\delta_{ij}dx^idx^j\ ,\quad i=1,\dots,d-1\ ,\quad
r\in[r_0,\infty)\ .
\end{equation}
Assuming that $g\to r^2$ and $c\to 1$ for large $r$ and $g$ has a (double) single zero at (zero) finite temperature, that defines the horizon, we follow the procedure devised by Donos and Gauntlett \cite{donos2014} to compute the dc conductivity from the solution of the EOM's at the horizon.

We add a perturbation in $A_x$ linear in time,
while the axion and metric perturbations are independent of time,
\begin{equation*}
A\to A+(a_x(r)-Et)\dd x,\quad X^x \to X^x+\chi(r),\quad \dd s^2\to \dd s^2+
2r^2 h_{tx}(r)\dd t\dd x +2r^2 h_{rx}(r)\dd r \dd x\ .
\end{equation*}
Maxwell's equation for $a_x$ is:
\begin{equation}\label{eq:ax_eom}
\partial_r\left[ Y\sqrt{-g}g^{rr}(g^{tx}F_{rt}+g^{xx}F_{rx}) \right]=0\ ,
\end{equation}
which leads to the radially conserved quantity:
\begin{equation}\label{eq:J}
J=-Yr^{d-3}c^{d-3\over 2}g a_x' - h_{tx}{\rho\over c}\ .
\end{equation}
This conserved quantity is evaluated at the horizon where $h_{tx}$ and $a_x$
are obtained as we discuss below.

The perturbation on the gauge field close to the horizon is obtained from
eq.
(\ref{eq:J}) by choosing $J$ such that $a_x$ is ingoing in the horizon:
\begin{equation}\label{eq:axp}
a_x'\sim -{E\over g}\implies a_x\sim -E v\ ,
\end{equation}
where $v$ is the ingoing Eddington-Finkelstein coordinate $v = t + r_*$, given
in terms of the tortoise coordinate $dr_*= {dr\over g}$. 

Eq. (\ref{eq:axp}) gives
the first term inside the parenthesis of eq. (\ref{eq:J}). To obtain the second
term, we combine the $(xt)$ and $(xx)$ Einstein's equations. Since we will evaluate them at the horizon, we will only write down explicitly the non-zero terms
after taking the limit $r\to r_0$.

For clarity we write down Einstein's equations only for $d = 3$:

\begin{equation}\label{eq:EEs_Z}
\begin{split}
ZG_{ab}&={1\over2}T_{ab}+\half{\cal Z}_{ab}\ ,\\
T_{ab}&=Y\left(F_a^cF_{bc}-{1\over4}g_{ab}F^2\right)-(2\Lambda+V)g_{ab}+\sum_I\nabla_aX^I
\nabla_b X^I \left( -\dot{V} - {\dot{Y}\over 4}F^2 + \dot{Z}R \right)\ , \\
{\cal Z}_{ab}&=2(\nabla_a\nabla_b  - g_{ab}\nabla_c\nabla^c) Z\ , \\
\end{split}
\end{equation}
where the dot derivative stands for derivative with respect
to $TrX={1\over2}g^{ab} \sum\limits_I
\partial_aX^I\partial_bX^I$ and $R$ is the Ricci
scalar.\footnote{The full dynamical stability of the model
eq. (13) is beyond the scope of the paper. We thank Oriol
Pujol\`as and Matteo Baggioli for pointing out the
occurrence of third order time derivatives. These occur beyond the
linear analysis and have the potential to further restrict
the parameters for which the model is stable.}

The $(tx)$ Einstein's equation is,
\begin{equation}\label{eq:EExt}
\begin{split}
&{\cal O}(h_{tx}',h_{tx}'') + {r^2\over2}Zh_{tx} r^2 \left[ g''+g'\left( {2\over
	c}+{c'\over c} \right)  + \dots\right]= \\
&{r^2\over4}h_{tx}\left(-4\Lambda - 2V + Y a_t'^2 \right) + \half gY a_t a_x' +
\half r^2g \dot{Z} TrX'h_{tx}'- {r h_{tx} \over 2c } \left[2rcg' \dot{Z} TrX' +
\dots \right]\ , \\
\end{split}
\end{equation}
where the dots and the terms ${\cal O}(h_{tx}',h_{tx}'')$ are zero at the horizon,
the prime derivative is with respect to $r$ and $TrX'=\pa_rTrX$. The first two
terms on the right-hand side come from $T_{ab}$ and the last term from ${\cal
	Z}_{ab}$. In order to simplify eq. (\ref{eq:EExt}) we eliminate $c''(r)$ from
the $(tt)$ Einstein's equation and substitute it into the $(xx)$ Einstein's
equation. The result is given in eq.
(\ref{eq:EExx}) by specifying $G_{xx}$, $T_{xx}$ and ${\cal Z}_{xx}$
separately,
\begin{equation}\label{eq:EExx}
\begin{split}
ZG_{xx}&={r^2c\over2}Zg''+ {r^2 Z\over 4}\left({c'\over c}+{2\over r}\right)g'
+ {r^4Z\over 8}{4\Lambda +2V + Y a_t'^2\over k^2\Zp-r^2cZ
}+\dots \ ,\\
{\cal Z}_{xx}&={\alpha^2\Zp g'\over2}\left({c'\over c}+{2\over r}\right) + 
{\alpha^2r^2c\Zp\over 4}{4\Lambda + 2V + Y a_t'^2\over \alpha^2\Zp-r^2cZ}+\dots\ , \\
\half T_{xx}&={\alpha^2\dot{Z}\over2}g'' + 
{\alpha^2 \dot{Z}\over  2}\left({c'\over c}+{2\over r}\right)g' +
{r^2 Z\over 4}{r^2c(4\Lambda+2V-Y a_t'^2) + \alpha^2(-2\Vp+\Yp a_t'^2)
	\over \alpha^2\Zp-r^2cZ} \\
&+ {\Zp\over 4}{2\alpha^2\Vp + (2\alpha^2r^2cY-\alpha^4\Yp)a_t'^2 \over
	\alpha^2\Zp-r^2cZ}+\dots\ ,
\\
\end{split}
\end{equation}
where the dots vanish at the horizon. Combining eqs. (\ref{eq:EExx}) and
(\ref{eq:EExt}) allows to eliminate $h_{tx}$. Its value at the horizon
$r=r_0$, is used to calculate the dc conductivity from $\sigmadc=J/E$. Before we
do so, we define the expansions of the metric functions $g$ and $c$ close to the horizon as:

\be\label{eq:metric_expansion}
g~\sim g_1\left(1-{r_0\over r}\right)+\dots, \ c~\sim c_0 + c_1
\left(1-{r_0\over r}\right)+\dots\ .
\ee
Restoring arbitrary bulk dimensionality $d+1$, the temperature is  
\begin{equation}\label{eq:Temp}
T={1\over 4\pi }\  {c_0 r_0\over 2c_0+c_1 }\ 
{4\Lambda+2V(r_0)+Y(r_0) a_t'(r_0)^2\over {2\alpha^2\Zp(r_0)\over
r_0^2c_0}-(d-1)Z(r_0) } \ ,
\end{equation}
where we have used $TrX={\alpha^2\over r^2c}$ to simplify the denominator. The
dc conductivity, obtained from eqs. (\ref{eq:J}), (\ref{eq:axp}) and the value
of $h_{tx}$ at the horizon, calculated as indicated previously, are given by the
following compact expression,
\begin{equation}\label{eq:sigma}
\begin{split}
\sigmadc&=Y_0 r_0^{d-3}c_0^{d-3\over 2} + {\rho^2\over m_{eff}^2}\
,m_{eff}^2=2c_0^{\frac{d-1}{2}} r_0^{d-1}\left(T{\cal A}+{Z_0^2 {\cal B} + \Zp_0 {\cal
		C}}\right)\\
\end{split}
\end{equation}
where $T$ is the temperature eq. (\ref{eq:Temp}) and ${\cal A}$, ${\cal B}$
and ${\cal C}$ are given in eq. (\ref{eq:sigma_cont}). The term $a_t'={\rho\over
Y(r) r^{d-1} c(r)^{d-1\over 2}}$ is also evaluated at the horizon.
The subscripts $'0'$ in eqs. (\ref{eq:sigma}) and (\ref{eq:sigma_cont})
indicate the variable is evaluated at the horizon.

Eq. (\ref{eq:sigma}) suggests that even at zero temperature the conductivity
receives a correction given by the $\Zp_0^2{\cal C}$ term. We note that although this is a fully analytical expression for the conductivity the metric at the horizon may only be computed numerically. 
\begin{equation}\label{eq:sigma_cont}
\begin{split}
{\cal A}&={4\pi (2c_0+c_1)\over r_0^3 c_0^3} {{(d-1)(d-2)\over 4}r_0^4c_0^2Z_0^2
	- (d-2)\alpha^2r_0^2c_0Z_0\Zp_0 + 2\alpha^4\Zp_0^2\over 
	(d-1)Z_0-{2\alpha^2\Zp_0\over r_0^2c_0} 
}\ ,\\
{\cal B}&= {(d-1)c_0 \over 4\left[(d-1)Z_0-{2\alpha^2\Zp_0\over r_0^2c_0}
	\right]^2} \left[ (d-2)r_0^2c_0\big(4\Lambda+2V_0+Y_0 a_t'^2\big)
+2\alpha^2\big(2\Vp_0-\Yp_0 a_t'^2\big)\right]\ ,\\
{\cal C}&={\alpha^2\over r_0^2c_0^2\left[(d-1)Z_0-{2\alpha^2\Zp_0\over r_0^2c_0}
	\right]^2}
\Big\{ \alpha^2\Zp_0 \big( 4\Lambda+2V_0-Y_0 a_t'^2 \big) - \\
& \hspace{5cm} Z_0\Big[{d-1\over2}r_0^2c_0\big(12\Lambda+6V_0+Y_0 a_t'^2\big) + 
\alpha^2\big( 2\Vp_0-\Yp_0 a_t'^2 \big)
\Big] 
\Big\}\ . \\
\end{split}
\end{equation}
\subsection{Conductivity reduction induced by charge screening}\label{sec:charge_screening}

Our result for the dc conductivity generalizes those obtained previously from
the AdS $RN+$axions background ($Z=1$, $Y=1$ $V=TrX$) \cite{Andrade2014}, and
from the backgrounds studied in \cite{Baggioli2016,Gouteraux2016} with $Z=1$,
$Y=e^{-\kappa TrX}$ $V=TrX$, $\kappa>0$. In these models the Ricci scalar is not
coupled directly to the axion. However, the axion-dependent coupling $Y$ has a
crucial role in the dc conductivity.

For sufficiently large $\kappa$ and $\alpha$, and for low temperature, the dc conductivity increases with temperature, a behaviour previously referred to as \textit{insulating} \cite{Baggioli2016,Gouteraux2016}. We note that the physical reason for this behavior is not a smaller scattering time but the simple fact that $Y$ screens the charge at low temperature and consequently reduces the conductivity which is proportional to the charge. While the overall temperature dependence is similar to that expected in a system approaching an insulating state, it should be noted the conductivity is {\it always} finite so the system is metallic in all cases. Moreover, the scattering time, controlled by the parameter $\alpha$, in this model has the same temperature dependence than in $RN$+axions background for which no insulating behavior was observed. Very likely, a truly insulating behavior would lead to a qualitative change in the background something that is not observed in \cite{Baggioli2016,Gouteraux2016}.

In the following sections we identify a region of parameters in BD holography
where we have found similar features which are not induced by charge screening. However, we do not claim that our system is an insulator because the conductivity is never zero, even at zero temperature.

\subsection{The dc conductivity in BD axion backgrounds}
In this section we explore the effect of the BD-type coupling $Z$ on the
background and on the dc conductivity in $d=3$ boundary dimensions. 
More specifically, we break translational invariance by using an axion-dependent BD coupling $Z(TrX)$. 
First we consider the simpler case  $V=TrX={1\over d-1}\sum_I \nabla_\mu X^I\nabla^\mu X^I = 0$ in eq.
(\ref{eq:action}) and later study a more general BD-like model where $V=TrX \neq 0$ is also present. 
Initially we restrict our analysis to two space dimension. The dependence on dimensionality of our results is discussed in the last part of the section.

We note physically $Z(TrX)$ is an axion-dependent gravitational coupling constant that runs from the boundary to the horizon. 
For that reason we will refer to these axions coupled to the Ricci tensor as gravitational axions. The qualitative effect of this running in the holographic dimension, from weak to strong coupling, is less obvious than for $Y \neq 1$ or in the translational invariant case.

\subsubsection{Momentum relaxation with $Z\neq 1$, $Y=1$ and
$V=0$}\label{sec:axions2}

We start our analysis with the simpler case of no axion potential and trivial
coupling to the Maxwell tensor,
\be\label{eq:couplings_simpler}
\begin{split}
	&Z=e^{\lambda TrX}\ ,\quad \Zp=\lambda Z\ ,\quad TrX={\alpha^2\over r^2 c}\ ,\\
	& V=0\ ,\quad \qquad \Vp=0\ ,\\
	& Y=1\ , \quad \qquad \Yp=0\ ,\\
\end{split}
\ee 

Although, {\it a priori}, $\lambda$ and $\alpha$ are independent parameters, it
is easy to see from eq. (\ref{eq:couplings_simpler}) that these parameters
appear only in $Z$ as a single parameter
$\lambda_{eff}=\lambda\alpha^2$. Translational symmetry is
broken for both $\lambda_{eff}>0$ and $\lambda_{eff}<0$. In the
first case $c(r_0)<1$ while in the second $c(r_0)>1$.
However, for $\lambda<0$ the squared 'effective mass' $m_{eff}^2$ in eq.
(\ref{eq:sigma}) is negative. Therefore, $\lambda_{eff}$ should be constrained
to positive values.\footnote{As we will demonstrate later in a similar
background we have observed that $\lambda<0$ also leads to the violation of the null energy condition.
Therefore, we restrict to positive $\lambda_{eff}$.}

At high temperature, the background tends to AdS $RN+$axions \cite{Andrade2014} ($Z=1$, $Y=1$,
$V=TrX$): it has a similar blackening factor and $c\to1$ for all $r$. The
effect of the coupling $Z$ in the background is more evident at low
temperatures where $c(r)$ has a stronger dependence on the radial dimension.

Regarding the dc conductivity, this model has qualitatively similar properties
to that of the $RN+$axions. For example, for the allowed range of parameters, the dc
conductivity increases as the temperature decreases and $\sigmadc >Y_0=1$. Moreover, it is known
\cite{Andrade2014} that for $RN+$axions in four bulk dimensions, this condition is
$\sigmadc>r_0^{d-3}$. In the model eq. (\ref{eq:couplings_simpler}) the
condition is $\sigmadc>\left(r_0^2c_0\right)^{d-3\over2}$, where
$c(r_0)=c_0$. 

It would be interesting to compare the dc
conductivity for a fixed scattering time in the model of eq.
(\ref{eq:couplings_simpler}) with the well-studied AdS $RN+$axions model. However,
it is clear that both models display similar features since the respective actions are related by a conformal transformation.
As was explained in detail in sec. \ref{sec:conf_transf} for the translational
invariant case, under a conformal transformation, the action of eq.
(\ref{eq:action}) can be transformed into an action with $Z=1$. The change of the
Ricci scalar under this transformation involves additional terms which depend on $TrX$ and contain the usual kinetic term proportional to $\sum_I
\nabla_\mu X^I\nabla^\mu X^I$, $X^I=\alpha x^I$. \\
Moreover, as mentioned in sec. (\ref{sec:conf_transf}), in $d=3$ dimensions
electromagnetism is conformal, therefore, the conformal transformation does not
change the coupling of the $F^2$ term in the action, see eq.
(\ref{eq:coup_conf_transf}).
This is consistent with the fact that for the choice of couplings given in eq.
(\ref{eq:couplings_simpler}), the zero temperature dc conductivity is larger
than one, as in the $RN+$axions model \cite{Andrade2014}. In higher dimensions,
however, it is expected the model specified by eq. (\ref{eq:couplings_simpler})
will yield $\sigmadc<1$ in some range of parameters. Indeed we will see that this is the case in
sec. \ref{sec:higher_dim}.

In the next section we study a more general model with a finite potential ($V$) and non-trivial BD
($Z$) and Maxwell ($Y$) couplings in four bulk dimensions.

\subsubsection{Momentum relaxation with $Z\neq 1$, $Y\neq 1$ and $V\neq
0$}\label{sec:axions3}

Before showing the results for the dc conductivity, it is illuminating to
comment the general features of the gravitational background given in eq.
(\ref{eq:metric}) for the following choice of couplings,
\be\label{eq:couplings}
\begin{split}
	Z&=e^{\lambda TrX}\ ,  \qquad \qquad \Zp=\lambda Z\ ,\\
	 V&=TrX={\alpha^2\over r^2 c(r)}\ ,\ \Vp=1\ ,\\
	Y&=e^{-\kappa TrX}\ , \ \quad \qquad \Yp=-\kappa Y\ ,\\
\end{split}
\ee 
where $\alpha,\kappa >0$ and $\lambda$ is real.

The extremal charge density is: 
\begin{equation}\label{eq:charge_density}
\rho_e=r_0^2c_0\sqrt{Y(-2\Lambda-V)}\ .
\end{equation}
It is clear that $\rho_e$ decreases as $Y$ decreases, which for the choice of
eq. (\ref{eq:couplings}) corresponds to increasing $\alpha$ and $\kappa$.
Similarly, for smaller $c(r_0)=c_0$ the extremal charge density is smaller.

We now comment on the allowed range of the BD coupling parameter $\lambda$
according to the properties of the background.
Similarly to sec. \ref{sec:axions2}, $\lambda>0$ ($Z>1$) is allowed, in which
case the $g/r^2$ and $c(r)$ increase monotonically towards the boundary.
Contrary to the model of sec. \ref{sec:axions2}, $\lambda=0$ is also allowed due
to the presence $V$, which breaks translational invariance. In this case the BD
coupling is trivial $Z=1$ and has been studied previously
\cite{Baggioli2016,Gouteraux2016}.

Moreover, we also find backgrounds satisfying all boundary conditions for
$\lambda<0$ ($Z<1$). However, in this case $g/r^2$ does not increase
monotonically towards the boundary; there exists a point inside the bulk where
the derivative of $g/r^2$ vanishes. This suggests the odd feature that the
background displays a repulsive behavior between this point and the
boundary, which may violate energy null condition.\footnote{We thank Roberto Emparan
for discussion and suggestions on this matter.} In \cite{Santos2007}
the energy conditions in theories of gravity different from Einstein's gravity
have been re-derived from Raychaudhury’s equation and imposing gravity to be
attractive. For the action given in eq. (\ref{eq:action}) the null energy condition reduces to
\begin{equation}\label{eq:NEC}
{1\over Z}(T_{ab}+{\cal Z}_{ab})n^a n^b\geq 0\ ,
\end{equation}
for all null vectors $n^a$ and where $T_{ab}$ and ${\cal Z}_{ab}$ are defined
in eq. (\ref{eq:EEs_Z}).
We have found that for $\lambda<0$ the background violates the null energy condition eq. (\ref{eq:NEC}). This is easily seen by
expressing the background eq.
(\ref{eq:metric}) in the variable $u=r_0/r$ and plugging the following null
vector:
$N^t=1/\sqrt{g}$, $N^r=\sqrt{g}$, $N^i=0$ in eq. (\ref{eq:NEC}), which reduces
to
\begin{equation}\label{eq:null_energy}
(T_{ab}+{\cal Z}_{ab})N^a N^b\propto \ddot{Z} TrX'^2 +
\dot{Z} TrX'' \propto c'(u)^2-2c(u)c''(u)\geq 0\ .
\end{equation}
For $\lambda\geq 0$, $c''(u)\leq0$ and the null energy condition for this null
vector is satisfied, however we have observed that for any $\lambda< 0$ this
condition is violated.

Regarding the metric function $c(r)$,
as the temperature increases, it becomes almost independent of the holographic
coordinate $c(r) \approx c(r_0)=c_0\to 1$. The
blackening function is also modified in such a way that the geometry approaches
that of AdS $RN+$axions.
In the allowed region: $\lambda\geq 0$, the horizon value of $c$ satisfies
$c_0\leq1$ and $c_0$ decreases for larger $\lambda$.
We have already observed this behavior in the model of Sec. \ref{sec:axions2}
($Y=1$ and $V=0$). On the other hand, in the forbidden region $\lambda<0$,
$c(r_0)=c_0>1$ and $c_0$ increases for smaller $\lambda$. We note that,
at low temperatures the spatial metric functions
$g_{ii}=r^2c(r)$ could, in principle, be better understood in terms of Lifshitz
and  hyperscaling violation anomalous exponents, similarly to EMD theories 
\cite{kiritsis2015}.
While we do not rule out the behavior of $g_{ii}$ close to the horizon may
actually be cast using various anomalous exponents, we have not been able to
re-express the metric at low temperatures
using a single anomalous exponent.

Finally, in this model, contrary to that of sec. \ref{sec:axions2},
$\lambda$ and $\alpha$ are independent parameters. In the presence of
$V=TrX={\alpha^2\over r_0^2c_0}$, the parameter $\alpha$ appears 
independently of the parameter $\lambda$ in the action and the equations of
motion. Therefore, it is expected that these two cannot be relabelled into a
single parameter. For more explicit results regarding the background for
different choices of the parameters, see the appendix \ref{appZYV}.

\subsubsection*{The dc conductivity}
\begin{figure}[b]
	\includegraphics[scale=0.75,center]{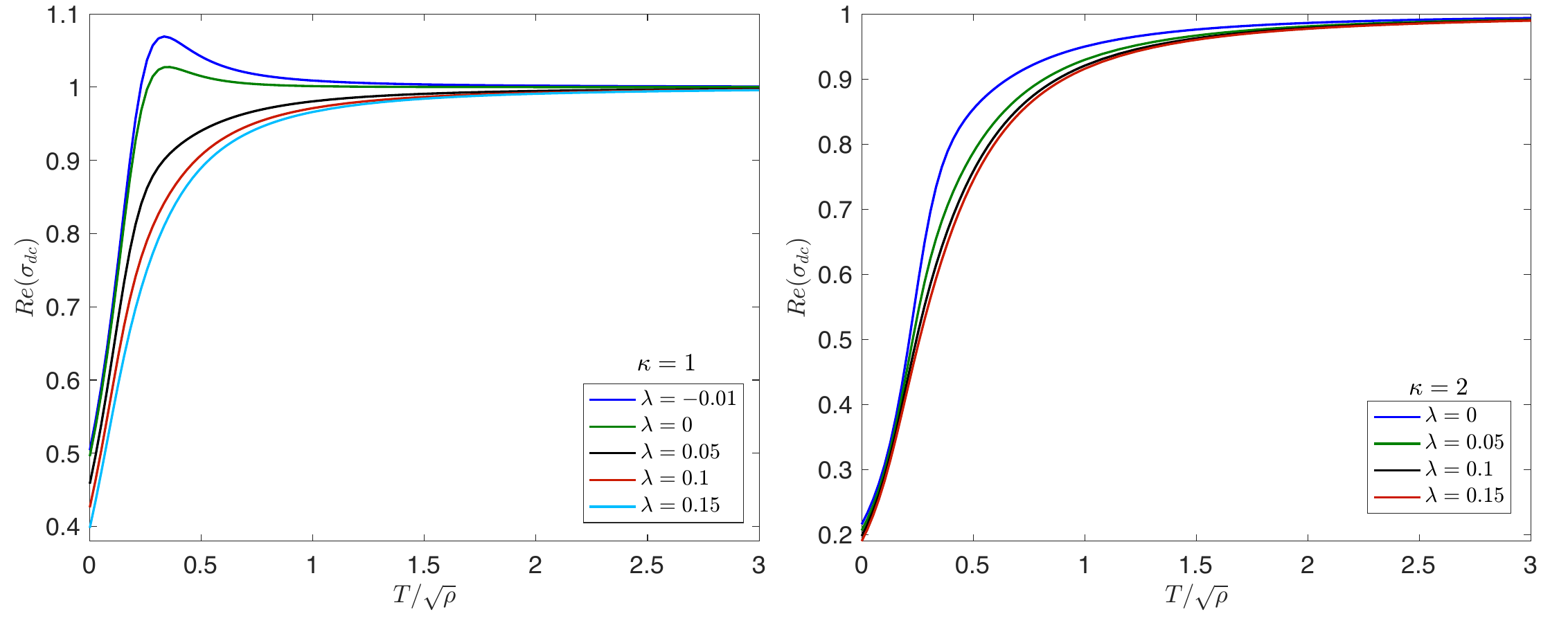}
	\vspace{-10mm}
	\caption{Temperature dependence of the dc conductivity eq. (\ref{eq:sigma}) for fixed
	axion parameter $\alpha=1$ and charge density $\rho=1$.
	The charge screening parameter $\kappa$ and the BD-coupling parameter $\lambda$ are
	indicated in the plots. The effect of increasing $\lambda$ for fixed $\kappa$ is similar to
	increase $\kappa$ for fixed $\lambda$. The case $\lambda<0$ suggests
	that the effect of a weaker gravitational coupling $Z<1$ on the dc conductivity
	is to weaken momentum dissipation. However, we stress this limit violates the null
	energy condition eq.
	(\ref{eq:null_energy}) and should be excluded.}\label{fig:dcvT_Kap1_Kap2}
\end{figure}

\begin{figure}[t]
	\center
	\hspace{-5mm}
	\includegraphics[scale=0.6,clip]{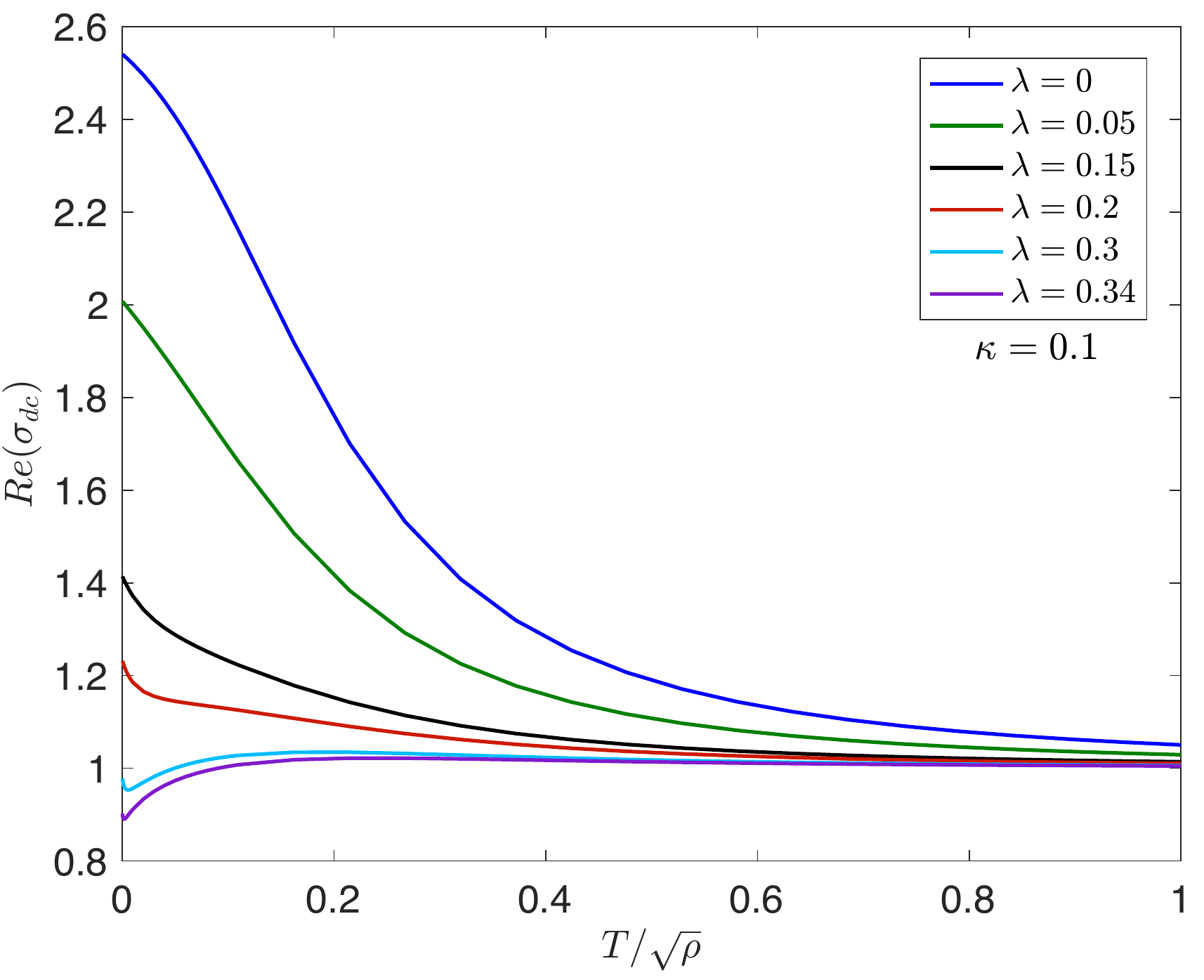}
	\vspace{-4mm}
	\caption{Temperature dependence of the dc conductivity eq. (\ref{eq:sigma}) for fixed
	axion parameter $\alpha=1$ and charge density $\rho=1$.
	The charge screening parameter $\kappa$ and the BD-coupling parameter
	$\lambda$ are indicated in the plot. The effect of the BD
	parameter is similar to the charge screening parameter.
	Increasing $\lambda$ yields a smaller $r_0^2c_0$; as a
	consequence the first term in eq. (\ref{eq:sigma}): $Y_0={\rm
	exp}\left[-\kappa{\alpha^2/( r_0^2c_0)}\right]$ decreases. The bound of \cite{Grozdanov2015} is violated for large $\lambda$ even for weak charge screening.}\label{fig:dcvT_Kap01}
\end{figure}

We depict in
Figs. \ref{fig:dcvT_Kap1_Kap2} and  \ref{fig:dcvT_Kap01} the dc
conductivity eq.
(\ref{eq:sigma}), in two space dimensions as a function of temperature for a
wide range of the BD parameter $\lambda$ and the charge screening parameter
$\kappa$. The effect of $\lambda$ and $\kappa$ is very similar: both control the strength of
momentum dissipation. 

In Fig. \ref{fig:dcvT_Kap1_Kap2} we observe that the increase of either the
charge screening or the effective gravitational coupling ($Z>1$) yields a lower dc conductivity
especially for low temperatures and sufficiently large values of
$\lambda > 0,\kappa$. We note that that in this range of parameters the conductivity is below the bound only because of charge screening. 

However, in Fig.
\ref{fig:dcvT_Kap01} we observe that, even though the BD parameter does not
appear explicitly in the Maxwell coupling $Y$, its effect is to {\it
renormalize} the charge screening parameter $\kappa$ through the change in the
geometry. More explicitly, through the value of the metric function $c(r)$ at
the horizon: $c_0$. This is possible even for small $\kappa$. As mentioned
before, increasing $\lambda$ leads to a smaller $c_0$ and BD coupling which
manifests as stronger momentum dissipation.
This is a quite interesting and unexpected feature of the model.
For instance, the bound in the conductivity of \cite{Grozdanov2015} is violated,
even for very weak charge screening $\kappa = 0.1$ provided that momentum dissipation by
gravitational axions is strong enough $\lambda \geq 0.3$. The change in the temperature dependence of the dc conductivity at low temperature for different values of the parameters is not caused by 
charge screening but by the effective running of the gravitational coupling. As anticipated in sec. \ref{sec:charge_screening}, a decrease in the conductivity for low temperatures sometimes occur in systems that approach an insulating transition. However, in our case the conductivity, though substantially suppressed, it never vanishes. Therefore the model we study is never an insulator.

Notice that in the left plot of Fig.
\ref{fig:dcvT_Kap1_Kap2} we have included a case in the forbidden range of the
BD parameter: $\lambda<0$, which corresponds to a BD coupling satisfying $Z<1$
and $\Zp=\partial_{TrX}Z<0$. We have included this value only to tentatively
suggest that the effect of a weaker gravitational interaction could be to
effectively reduce the strength of momentum dissipation. 
Moreover, we have
observed that the effective mass in eq. (\ref{eq:sigma}) becomes negative for some
$\lambda_m<0$, which depends on the rest of the parameters. A negative
effective mass has been linked to instabilities of the theory
\cite{Baggioli2016}. However, we emphasize that, even for  our choice of the BD
coupling $Z$ there is a region $\lambda_m<\lambda<0$ in which the effective mass
is positive but the null energy condition  eq.(\ref{eq:null_energy}) is
violated.

\section{BD holography in higher dimensions}\label{sec:higher_dim}
So far we have restricted our analysis to $d+1=4$ bulk dimensions.
Here we briefly discuss the most salient features of higher dimensional
backgrounds. The motivation to study $d>3$ is to observe the explicit effect of the running of the gravitational constant associated to the extra factor $g_{xx}^{d-3\over 2}=r_0^{d-3}c_0^{d-3\over 2}$ in the first term of the conductivity eq.
(\ref{eq:sigma}),
\begin{equation}\label{eq:sigma_term1}
\sigmadc=Y_0 r_0^{d-3}c_0^{d-3\over 2} + \dots\ .
\end{equation}
The presence of this term is not exclusive
to the BD model. Indeed, in EMD models where translational invariance is broken
by axion fields, the same factor is present, \cite{Gouteraux2014}. We note
however that, at least in EMD theories where the axions and the dilaton are
coupled {\it minimally} through a dilaton-dependent coupling constant,
\cite{Gouteraux2014,kiritsis2015,garcia2015a}, the metric function $c$ is
trivial, $c=1$, \footnote{An additional difference in the background is that in the
metric ansatz given in eq. (\ref{eq:metric}), $g_{tt}=-1/g_{rr}$, which is
not the case in a EMD plus axion theory, \cite{kiritsis2015,garcia2015a}.}.
Therefore, in EMD-axion theories , $c_0 = 1$. 

\begin{figure}[t]
	\includegraphics[scale=1.07,center]{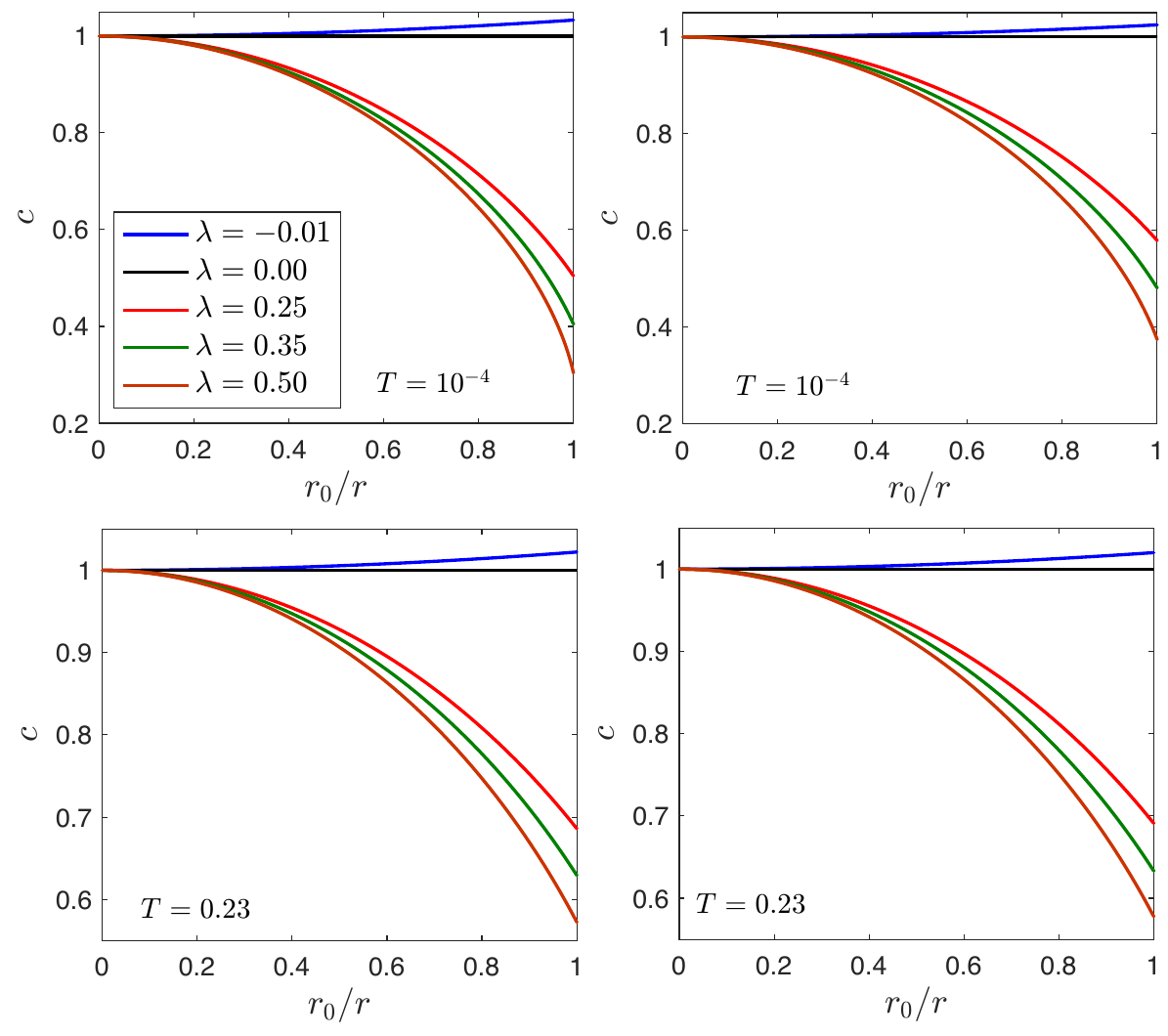}
	\vspace{-13mm}
	\caption{Metric function $c$ , eq. (\ref{eq:metric}), in $d+1=5$ (left column) 
	and $d+1=6$ (right column) bulk dimensions for the model given in eq.
(\ref{eq:couplings}). The
	temperature is indicated in the plots and the charge density is $\rho=1$.
	 The charge screening parameter $\kappa=1$ and the axion parameter $\alpha=1$. 
	 The BD-coupling parameter $\lambda$ is indicated in the legend, which refers
	 to all figures. While the boundary conditions for $\lambda<0$ are satisfied,
	 this case leads to violation of the null energy condition.}
	\label{fig:c_D5_6}
\end{figure}

We now discuss the two models of Secs. \ref{sec:axions2} and \ref{sec:axions3}
for $d=4,5$ boundary spacetime dimensions. In Fig. \ref{fig:c_D5_6} we plot
the metric function $c(r)$ used in the ansatz eq.
(\ref{eq:metric}), in the model of eq.
(\ref{eq:couplings}). In the absence of $V(TrX)$, eq.
(\ref{eq:couplings_simpler}), the background has similar features.

The results, depicted in Fig. \ref{fig:c_D5_6}, indicate that increasing dimensionality decreases the curvature of the metric function $c$. This is
more easily seen at low temperature (top row), where $c_0=c(r=r_0)$ increases
for $\lambda>0$ and decreases for $\lambda<0$.
Though not shown in the figure, a similar effect is also observed in the blackening function.
This is a manifestation of the large-dimensionality limit, \cite{emparan2013,garcia2015b} where the shape of $g$ and $c$ is such that the gravitational effects are
stronger closer to the horizon but weaker far from it.

 Dimensionality effects on the dc conductivity eq.
(\ref{eq:sigma_term1}) are directly related to the dependence of $c_0$ and $r_0$ on the dimension. The quantities with tilde are in $\tilde
d$ dimensions and those without tilde in $d$ dimensions. If $\tilde d>d$, we observe that:
\begin{itemize}
\item For $\lambda>0$ and low (high) temperature:  $\tilde c_0> (<) c_0$ and $\tilde r_0 < (>) r_0$.

\item For $\lambda<0$ (forbidden by the null energy condition) and low (high) temperature: 
$\tilde c_0< (>) c_0$ and $\tilde r_0> (<) r_0$.
\end{itemize}
For the temperature range studied, $\tilde r_0^2\tilde c_0 < 
r_0^2c_0$.  Moreover, for low temperature $r_0^2c_0<1$
but for large temperature $r_0^2c_0>1$. Therefore, for low temperature one
expects the suppression of the dc conductivity to be smaller for larger dimensionality.

\begin{figure}[H]
	\includegraphics[scale=0.76,center]{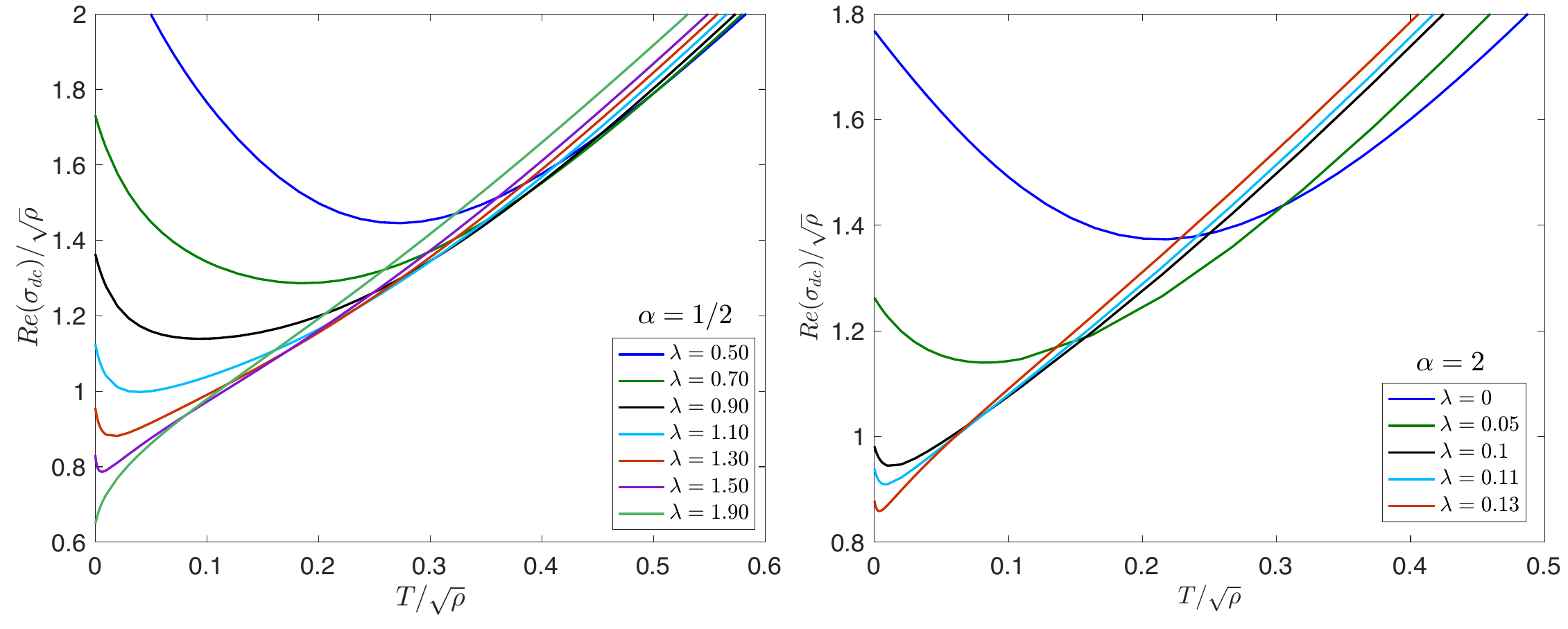}
	\vspace{-14mm}
	\caption{Zero frequency conductivity eq. (\ref{eq:sigma}) in $d+1=5$ bulk
	dimensions in a model with a minimally coupled Maxwell field, eq.
	(\ref{eq:couplingsD5}). When the BD coupling $\lambda$ is larger than some
positive value the dc conductivity at zero temperature is below $1$. This effect
is due to the background rather than to an axion-dependent Maxwell coupling as
in Fig. \ref{fig:dcvT_Kap1_Kap2}. The charge density is fixed $\rho=1$.}
	\label{fig:dcvT_Kap0_D5}
\end{figure}

We show in Fig. \ref{fig:dcvT_Kap0_D5} that the term $g_{xx}^{d-3\over2}$ in eq.
(\ref{eq:sigma_term1}) leads to a suppression of $\sigmadc$ at low temperature
in three space (boundary) dimensions ($d=4$). In order to isolate the effect of
the background we couple the Maxwell field minimally by setting $Y=1$, namely,
the couplings used are those given in eq. (\ref{eq:couplingsD5}).
\vspace{-5mm}

\be\label{eq:couplingsD5}
\begin{split}
	Z&=e^{\lambda TrX}\,,\   V=TrX={\alpha^2\over r^2 c(r)}\,,\ Y=1\,,\\    
	\Zp&=\lambda Z\,,\quad\ \ \Vp=1\,,\qquad\qquad\qquad \Yp=0\,.\\
\end{split}
\ee 

\begin{figure}[H]
	\includegraphics[scale=0.6,center]{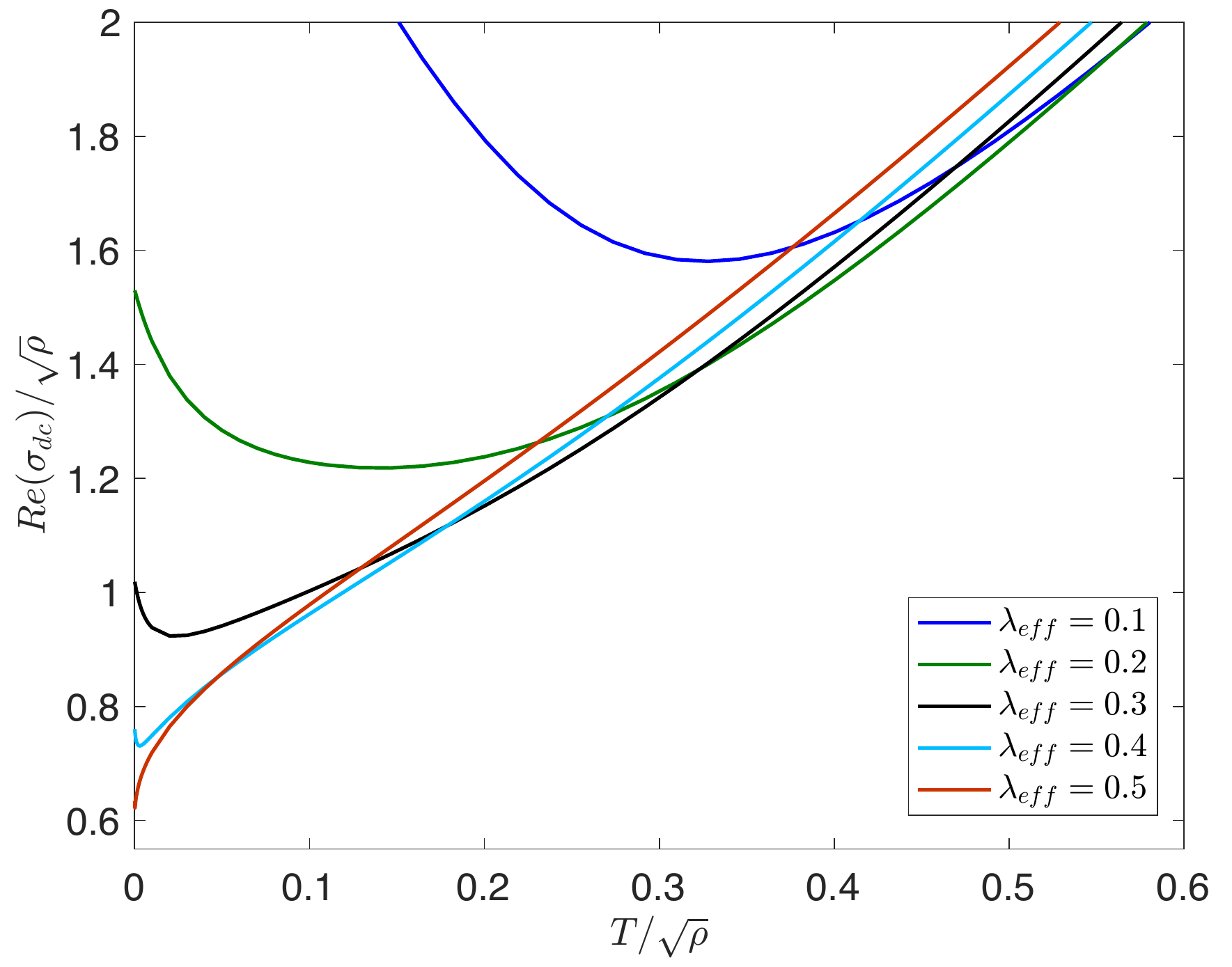}
	\vspace{-12mm}
	\caption{Zero frequency conductivity eq. (\ref{eq:sigma}) in $d+1=5$ bulk
	dimensions in a model with a minimally coupled Maxwell field and $V(TrX)=0$,
	eq.
	(\ref{eq:couplings_simpler}). Similar behavior as in Fig.
	\ref{fig:dcvT_Kap0_D5} is observed despite the absence of the usual 
	kinetic term in the action ($V=0$). Again, the effective change in the
	gravitational interaction, through the BD coupling $Z R$, parametrised by $\lambda_{\rm eff}= \alpha^2 \lambda$, allows to decrease
	the dc conductivity despite the absence of charge screening $Y=1$.
	The charge density is fixed $\rho=1$.}
	\label{fig:dcvT_Kap0_D5_noV}
\end{figure}

A similar effect, depicted in Fig. \ref{fig:dcvT_Kap0_D5_noV}, is observed even in the absence of $V(TrX)$ in the action (\ref{eq:action}), namely, we choose the couplings of eq.
(\ref{eq:couplings_simpler}). As was explained in Sec. \ref{sec:axions2}, in the
absence of $V$, there is a single parameter that controls momentum dissipation
$\lambda_{eff}=\lambda \alpha^2>0$. In summary, our results suggest that for higher dimensions gravitational effects, including those of the gravitational axions are suppressed except close to the horizon. As a consequence, the conductivity is closer to the RN limit for high temperature. However close to zero temperature, where gravitational effects are still important, the dc conductivity is heavily suppressed for strong momentum relaxation induced by the gravitational axion only.\footnote{Although, as the case with charge screening of sec. \ref{sec:axions3}, the conductivity never vanishes.} Therefore, also in this case, the background induces a important suppression of the conductivity without the need of any external source of charge screening. Obviously the system studied in the paper is always a metal, however it would be interesting to explore whether there are other backgrounds within BD holography that reproduce genuine insulating features in the dual field theory.

\section{Optical conductivity in BD holography with momentum relaxation}
We continue our analysis of transport properties of BD holography by investigating the optical conductivity. 
We focus first in the low frequency scaling of the real part of the conductivity. We found that the conductivity grows linearly with the frequency for low temperatures but strong momentum relaxation. In the second part of the section we show that BD holography, even if combined with other sources of momentum relaxation, does
not reproduce the intermediate-frequency scaling of the absolute value and argument of the optical conductivity observed in cuprates.

\subsection{Low-frequency behavior of the conductivity}
In the context of massive gravity the equation for the perturbation leading to
the optical conductivity at extremality has been solved analytically for low
frequencies by using the method of matched asymptotic expansions
\cite{davison2013}.
In \cite{Andrade2014} it was shown that, in the previous model, the dc
conductivity is equivalent to that obtained in the $RN+$axions model, upon a convenient
identification of the parameters. Using the method of matched asymptotic
expansions we have observed (not shown) that, as in massive gravity, the
low-frequency behavior of the optical conductivity of the extremal $RN+$axions
model is also linear in frequency with an always negative slope, which is
consistent with Drude physics.

\begin{figure}[h]
	\center
	\includegraphics[scale=1.4,clip]{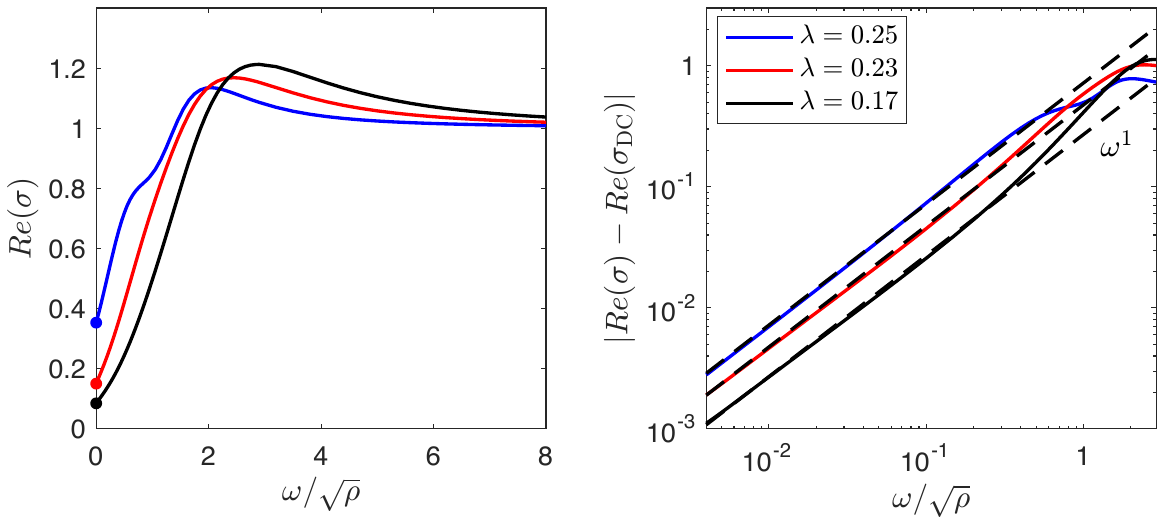}
	\vspace{-10mm}
\caption{Zero temperature [extremal background of eqs. (\ref{eq:action}),
(\ref{eq:couplings})] optical conductivity for strong
breaking of translational symmetry.
The charge screening parameter is $\kappa=1$ and the axion parameter
is $\alpha=1$ (blue lines), $\alpha= 1.5$ (red lines) and $\alpha=2$ (black
continuous lines). The BD-coupling parameter $\lambda$, given in the legend, is
fixed close to the maximum value allowed by the background boundary conditions.
The extremal charge density is $\rho=1$. The dashed black lines correspond to
the linear scaling $\beta
\omega$, where $\beta$ is fixed from the lowest frequency point of the
numerical data.
}\label{fig:ac_T0}
\end{figure}

Although we have not been able to obtain analytical results of the low-frequency
scaling for arbitrary couplings $Z$ and $Y$, we observe numerically, see Fig.
\ref{fig:ac_T0}, the same linear scaling of the conductivity for small frequencies. Interestingly,
provided that momentum relaxation is strong enough, the slope of this linear
growth is positive, namely, the conductivity increases with the frequency. This
is not exclusive of BD holography, it is also observed for $\lambda = 0$  in the
limit of strong momentum relaxation induced by the axion coupled to the Maxwell
field.
This is an interesting feature which we are not aware to have been reported in holographic systems which do not include charge screening.

\begin{figure}[H]
	\center
	\includegraphics[scale=1.1,clip]{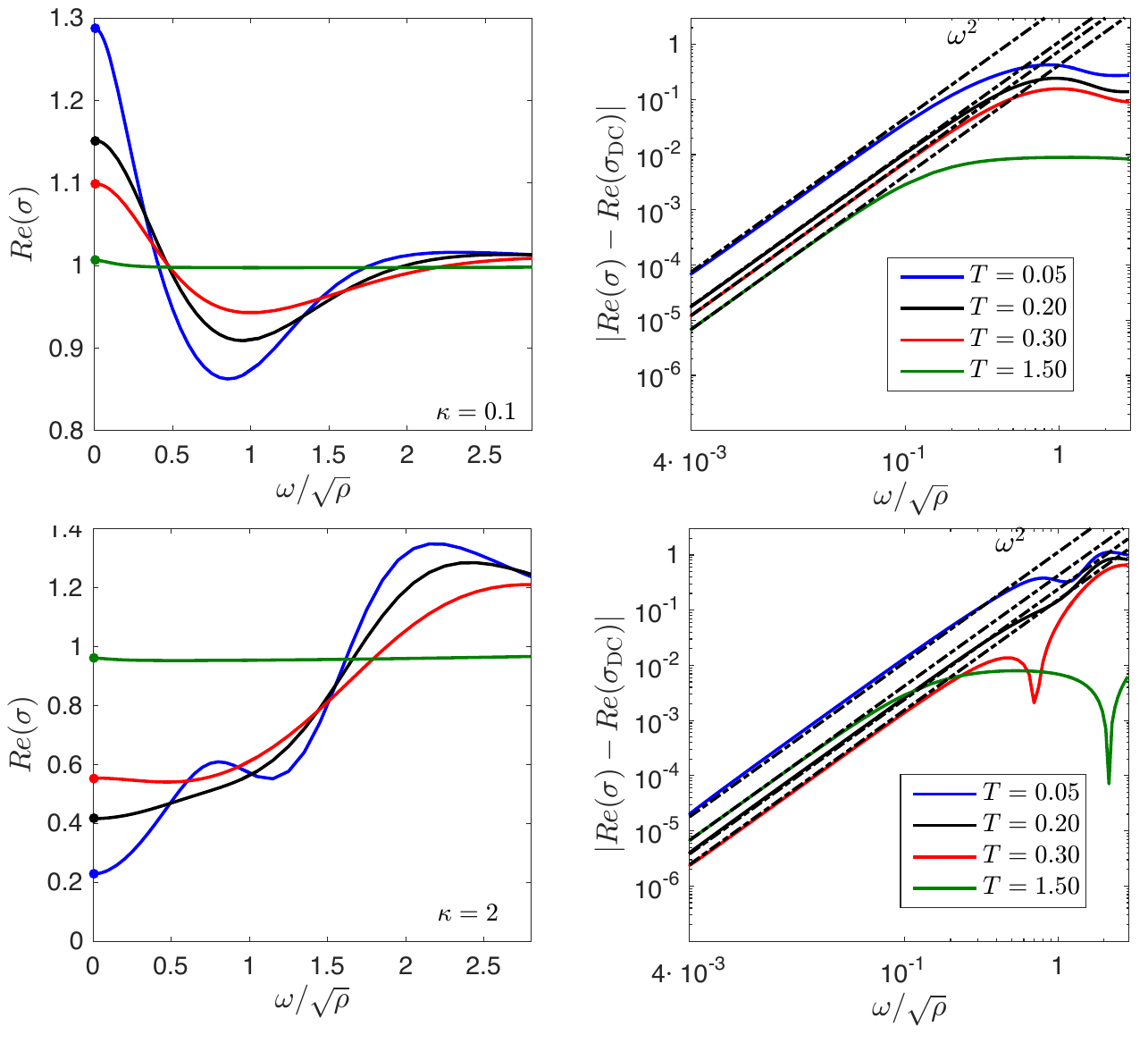}
	\vspace{-8mm}
\caption{Optical conductivity at finite temperature in the background given in
eqs.
(\ref{eq:action}) and (\ref{eq:couplings}).
Each line corresponds to a fixed temperature, indicated in the legend of the
right-hand side plots. The dashed-dotted black lines correspond to the
quadratic scaling and are fixed in the same way as in Fig. \ref{fig:ac_T0}.
As temperature decreases (blue lines) a slight disagreement is observed.
We expect for near extremal solutions the frequency scaling is given by eq.
(\ref{eq:freq_scaling}). The charge density $\rho=1$, $\alpha=1$, $\lambda=0.15$
and $\kappa$ is indicated in the left-hand side plots.}\label{fig:ac_Kap2_Kap01}
\end{figure}

At nonzero temperature, the numerical results of Fig. \ref{fig:ac_Kap2_Kap01}
show that the subleading term depends quadratically on the frequency. We
conclude therefore that for the general class of models with action eq.
(\ref{eq:action}), and for low frequency, the optical conductivity is
\begin{equation}\label{eq:freq_scaling}
{\rm Re}(\sigma)-\sigmadc=a \omega + b\omega^2+\dots
\end{equation}
where $b\to0$ for $T\to 0$ and, for $T\gg0$, both constants tend to zero, but
$a$ does it faster than $b$. In other words, at large temperature we have
observed a subleading contribution dominated by $\omega^2$ while in the limit of
zero temperature is proportional to $\omega^1$. 
As was mentioned above, for $Z=1$ and a minimally a coupled Maxwell field ($Y=1$),
the constants $a$ and $b$ are always negative, describing the broadening of the Drude peak.
For $Y\neq 1$ and both  $Z\neq 1$ and $Z=1$, $a$ is negative when $\sigmadc>1$
and positive when $\sigmadc<1$.

For $Y \neq 1$ and both at zero (not shown) and nonzero temperature (bottom left
plot in Fig. \ref{fig:ac_Kap2_Kap01}),  we have observed a range of parameters
for which the optical conductivity has a local maximum for relatively small
frequencies. In Fig.
\ref{fig:ac_Kap2_Kap01} we observe that in the high temperature limit the local peak
is smeared. A similar feature in the dc conductivity has been recently reported in \cite{Baggioli2016}. Similarly to the phenomenology observed in the temperature dependence 
of the dc conductivity, we believe that, in the range of parameters of Fig.
\ref{fig:ac_Kap2_Kap01}, this
intermediate peak is a consequence of charge screening induced by a non-trivial
axion-dependent Maxwell coupling and therefore it is not a precursor of insulating behavior. Indeed the conductivity is always finite. 

We have observed that a stronger gravitational interaction (larger BD coupling
$Z$) has a similar impact on the intermediate peak as increasing momentum
dissipation. As was mentioned in Sec.
\ref{sec:axions3},  in $d=3$ boundary dimensions this is a nontrivial effect: a
conformal transformation of the action eq. \ref{eq:action} with $d>3$
renormalizes the Maxwell coupling in an analogous way to the conformal
transformation of the BD model, see eq.
(\ref{eq:coup_conf_transf}). However, for $d=3$ the Maxwell coupling is
invariant under such transformation and one could expect therefore the Maxwell
coupling to be unchanged by a change in the BD coupling.
Nonetheless, from a holographic point of view, it is known that the observables
in the boundary theory are roughly determined by the gravitational background
(plus boundary values of bulk fields).
Therefore, at intermediate energy scales, and despite the fact the Maxwell
coupling is invariant under a conformal transformation for $d=3$, one should
expect the features of the BD background at intermediate length-scales to
determine the optical conductivity.

\subsection{Argument and modulus of the optical conductivity in BD gravity with momentum relaxation}
A well known property of the optical conductivity in most cuprates is that for
intermediate frequencies the module of the conductivity scales as
$\omega^{-2/3}$. It has been recently claimed \cite{horowitz2013} that a
holographic setup where momentum relaxation is introduced by a modulating
chemical potential share similar properties. However, we note that this
holographic setup does not reproduce another property of the optical conductivity
in cuprates: the argument of the conductivity is constant in the same range of frequencies.
In this section we study whether field theory duals of gravity models with different channels of momentum dissipation can reproduce these features of the optical conductivity in cuprates.

\begin{figure}[t]
	\center
	\includegraphics[scale=1.2,clip]{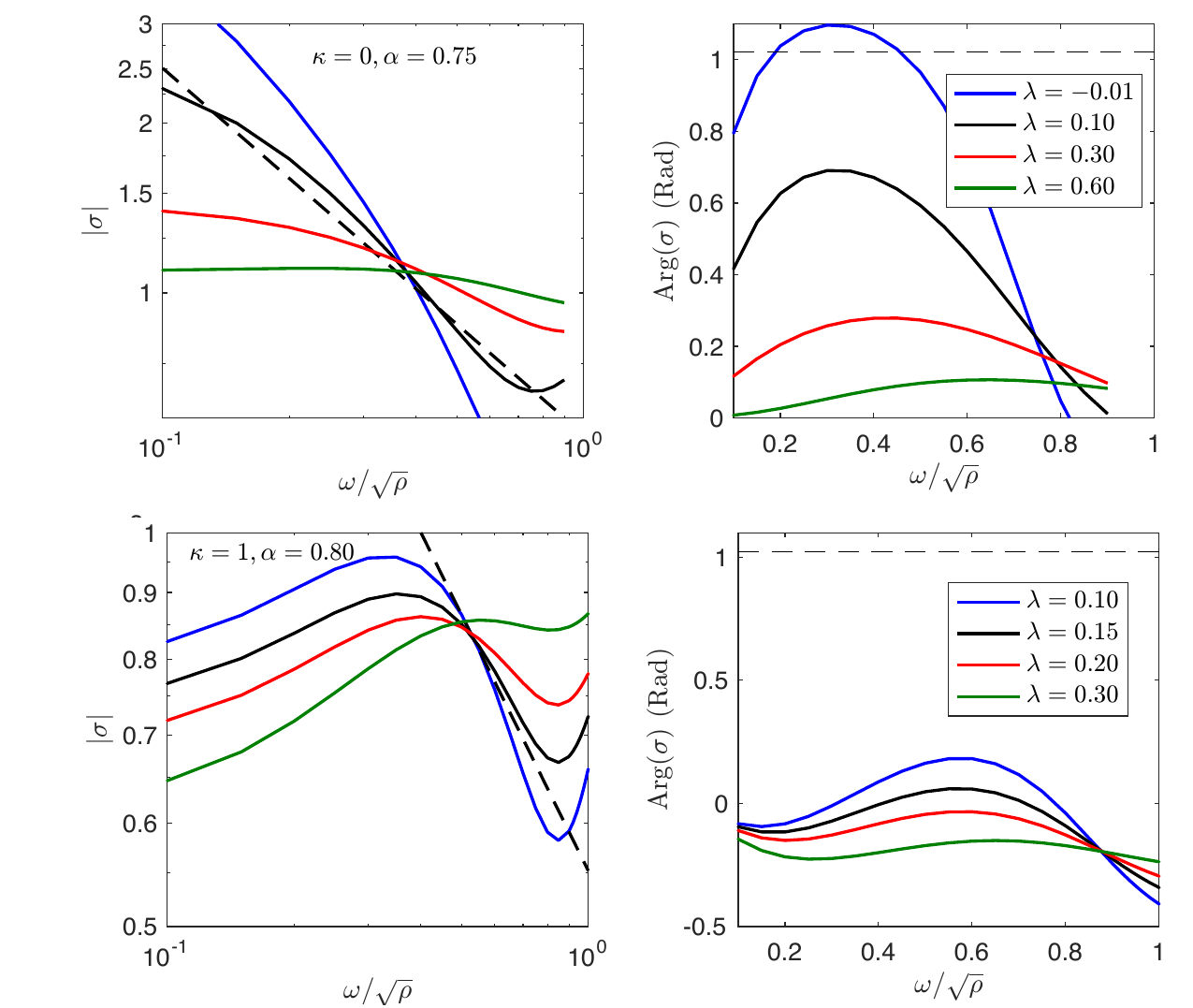}
	\vspace{-6mm}
	\caption{Absolute value and argument of the optical conductivity for
	different values of the parameters. The temperature and charge density
	are $T=0.05$ and $\rho=1$. The BD model with the couplings of
	eq. (\ref{eq:couplings}) does not describe the experimental behavior for
	intermediate frequencies observed for cuprates. The dashed lines are the
	exponent $\xi$, $|\sigma|=\omega^a$, $a=1.35-2=-0.65$, and
	$\mbox{arg}(\sigma)={\pi\over2}(2-1.35)=1.01$ Rad, according to experiments.}\label{fig:arg_cond}
\end{figure}

Results for different values of the parameters are depicted in
Fig. \ref{fig:arg_cond}.

Either the constant argument or the desired $2/3$ power-law decay can be observed for some values of the parameters. 
However, it is clear from our results that even by fine tuning all the available parameters we could not
reproduce both features for a single set of parameters.

\section{Ratio of shear viscosity and entropy density in BD with momentum relaxation}
In this section we study the ratio $\eta/s$ between the shear viscosity $\eta$
and the density of entropy $s$ for BD holography with momentum relaxation. In a
quantum field theory the viscosity is defined through the Kubo formula:
\begin{equation}
\eta=-\lim_{\omega\to0}{1\over \omega}\text{Im}G^R_{T^{xy}T^{xy}}(\omega,q=0)\ ,
\end{equation}
where $T^{xy}$ is the $xy$ component of the stress-energy tensor. In order to
compute the viscosity in the model of eq. (\ref{eq:action}) we use the membrane
paradigm in a similar way as it has been used in sec. \ref{sec:dc_BD} for the
calculation of the regular part of the dc conductivity. While
we restrict ourselves to numerical results we expect that, as shown in
\cite{Hartnoll2016}, it should be possible to derive  quasi-analytical
approximations at low and
large temperatures.\footnote{An analytical calculation in terms of the
background expansion close to the boundary is possible, however, the background
needs to be computed numerically.}
 \begin{figure}[b]
	\center
	\includegraphics[scale=1.1,clip]{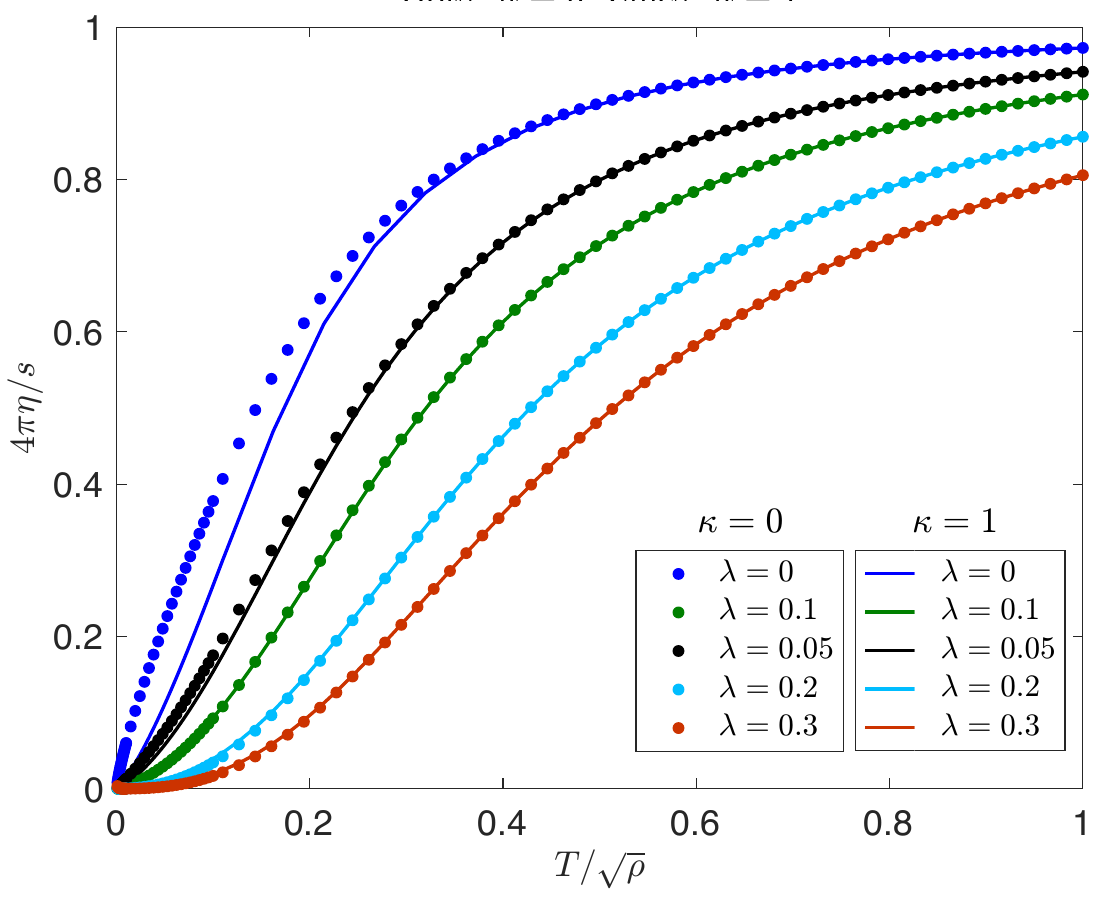}
	\vspace{-4mm}
	\caption{Shear viscosity to entropy density ratio ($\eta/s$) for the couplings of eq.
	(\ref{eq:couplings}) in the model defined in eqs. (\ref{eq:action}) and eq.
	(\ref{eq:couplings}).
	The axion parameter $\alpha=1$ and the charge density
	$\rho=1$.}\label{fig:KSS_V}
\end{figure}

\begin{figure}[h]
	\center
	\includegraphics[scale=1.1,clip]{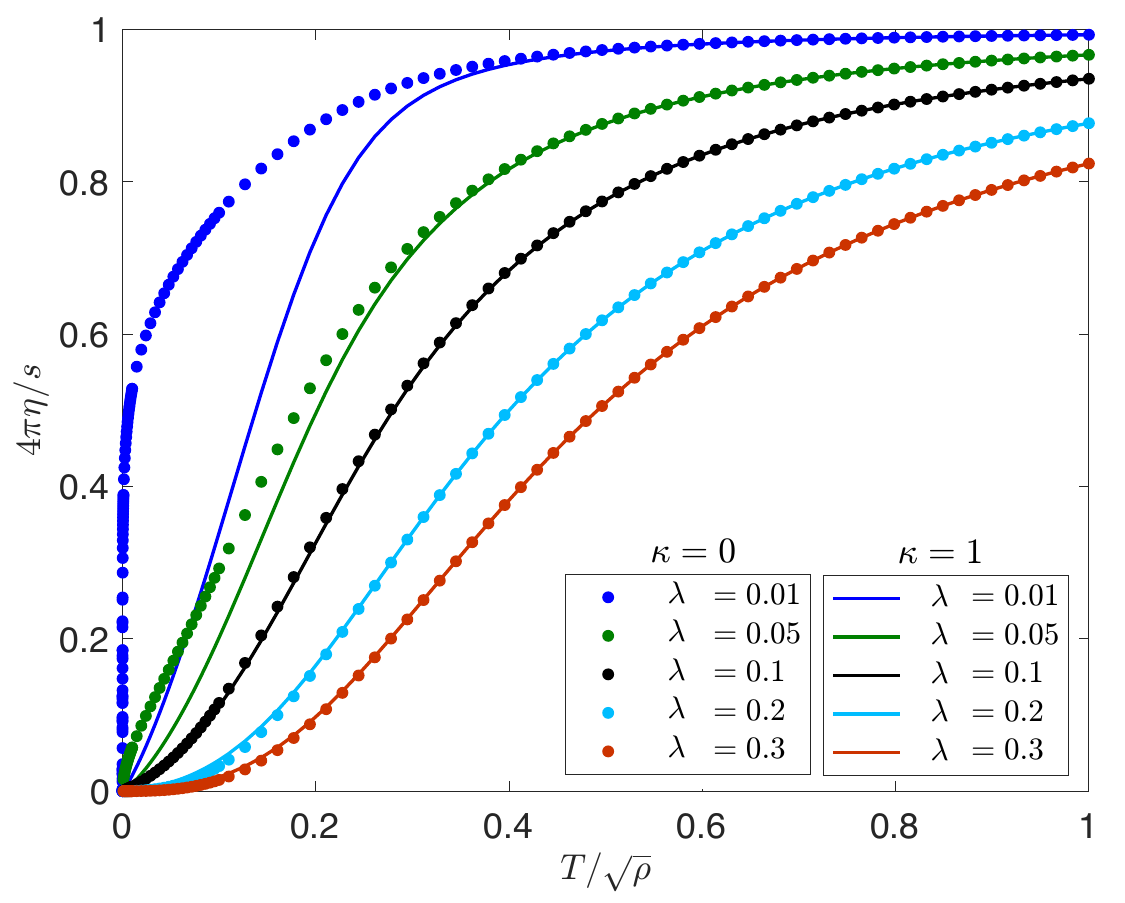}
	\vspace{-4mm}
	\caption{Shear viscosity to entropy density ratio ($\eta/s$) for the couplings of eq.
	(\ref{eq:couplings}) in the model defined in eqs. (\ref{eq:action}) and eq.
	(\ref{eq:couplings}) but taking $V=0$ and $\Vp=0$.
	The axion parameter $\alpha=1$ and the charge density
	$\rho=1$. As discussed in Sec. \ref{sec:axions2}, for $\kappa=0\implies Y=1$
	(dots) the theory has a single parameter that controls momentum
	dissipation, namely $\lambda_{eff}=\lambda \alpha^2$.}\label{fig:KSS_noV}
\end{figure}

Previously it has been reported that in the presence of
momentum relaxation \cite{Ge2015,Sadeghi2015,Burikham2016,Hartnoll2016,Ling2016} or anisotropy
\cite{Rebhan2012,Mamo2012,Finazzo2016}\footnote{We note that in these
models the effective mass of the graviton is nonzero. As shown in
\cite{Alberte2016} this is a necessary condition to observe violation of the KSS
bound. However, in theories without translational invariance and massless
graviton \cite{Alberte2016} the KSS bound remains valid. Our model falls in the
first class of theories.} the ratio is temperature dependent and in most
cases below the KSS bound.
We find similar results in the case of BD holography.
For our analysis we use the couplings of eq.
(\ref{eq:couplings}) with $V=TrX$ (Fig. \ref{fig:KSS_V}) and $V=0$ (Fig.
\ref{fig:KSS_noV}) which include, as a particular case, some of the previously
studied cases of AdS $RN+$axions \cite{Hartnoll2016,Ling2016}. In the full range of parameters we have explored, the ratio decreases with temperature
and is always below the KSS bound. It also decreases as the strength of momentum
relaxation increases, by any of the channels explored. It seems that it can be
made arbitrarily close to zero even for a finite momentum relaxation. We do not
have a clear understanding of the physical reasons behind this behavior however
we note that similar results have been observed \cite{giannakis2008} in the
context of the quark-gluon plasma with quenched impurities in the limit in which
the phenomenon of Anderson localization becomes important.

\section{Outlook and Conclusions}
The potential interest of BD backgrounds in holography is well beyond the
problems discussed in the paper.
For instance, the entanglement entropy depends explicitly on the gravity
coupling constant and it is also very sensitive to the strength of bulk
interactions.
Therefore we expect that the holographic entanglement entropy in inhomogeneous
BD backgrounds may reveal interesting features not found in previous holographic
duals. These features would not only be present to leading order
 (area of minimal surface) but also in the quantum correction
originated by the entanglement between the bulk and the minimal surface
\cite{Faulkner2013}. 
Another topic of potential interest is that of holographic
superconductivity. It is well known that the ratio between the order parameter
at zero temperature and the critical temperature, or the width of the coherence
peak are useful indicators of the strength of the interactions binding the
condensate. For the former, values much larger than the
Bardeen-Cooper-Schrieffer prediction, suggesting strong interactions, are
expected.
It would be interesting to investigate whether it is possible to tune this ratio
in BD backgrounds. That would be a smoking gun that the scalar in BD backgrounds
effectively controls the interactions in the bulk. Finally, we note that the
introduction of randomness in the scalar is qualitatively different from other
forms of disorder used in holography. It amounts to a random strength of the
gravitational interaction. Mobile charge introduced through the gauge field will
feel these random interactions not very differently from the way in which
electrons felt quench impurities. This is in stark contrast with the effect of a
random chemical potential, quite popular in holography, where the mobile
carriers are by construction randomly but homogeneously distributed through the
sample.
Coherence phenomena like Mott-Anderson localization could not be observed in
this setting. We plan to address some of these problems in the near future.

In summary, we have investigated the transport properties of strongly coupled
field theories whose gravity-dual is a Brans-Dicke action where gravity is
mediated by both a tensor, the graviton, and a scalar that depends on the radial
dimension. In the translational invariant limit we have computed analytically
several transport properties. The finite part of the dc conductivity $\sigma_Q$, expressed
in terms of thermodynamic quantities, is different from the universal prediction
for EMD backgrounds \cite{jain2010} however the shear viscosity ratio is still
given by the KSS bound. Similar results apply to other generalized f(R) gravity
backgrounds that can be mapped onto BD. The difference with EMD models is that
the entropy does not hold an area law as it also depends on the value of the
scalar at the horizon.

Momentum relaxation is induced by a gravitational axion, namely, the linear
coupling of the Ricci tensor and the axion. Following the procedure pioneered by
Donos and Gauntlett \cite{donos2014} we compute analytically the dc conductivity
as a function of the metric at the horizon which is evaluated numerically.
In $d+1=4$ bulk dimensions momentum relaxation by BD axions is qualitatively
similar to the results obtained by other mechanism of momentum relaxation
\cite{Andrade2014,Gouteraux2016,Baggioli2016} in the limit of strong charge
screening. Interestingly for strongly coupled gravitational axions, that induce
strong momentum relaxation, the conductivity bound \cite{Grozdanov2015} is
violated for any finite charge screening induced by the electromagnetic axion
\cite{Baggioli2016,Gouteraux2016}.
In higher spatial dimensions the dc
conductivity for sufficiently strong momentum relaxation decreases in the low
temperature limit. This suggests that the analogous conductivity bound is violated 
even if there is no coupling between the axion and the Maxwell field.
We have also computed numerically the optical conductivity in BD backgrounds
with momentum relaxation. For sufficiently strong breaking of translational
invariance, the conductivity grows linearly with the frequency in the limit of
small frequencies and very low temperatures though it remains finite for any temperature and frequency.
We have also evaluated numerically the modulus and
the argument of the optical conductivity for different momentum relaxation
channels in order to find out whether the phenomenology of this model is similar
to that of the cuprates for intermediate frequencies. Our results are not very
encouraging. For any value of the parameters we could not reproduce the
experimental results for both quantities simultaneously. Finally, we have shown
that the shear viscosity to entropy ratio decreases with temperature and the KSS
bound is violated by any strength of the momentum relaxation.

\acknowledgments 
A. R. B. thanks Roberto Emparan for illuminating
discussions. We also thank Matteo Baggioli and Oriol
Pujol\`as for useful comments on the manuscript. A. M. G.
acknowledges support from EPSRC, Grant No. EP/I004637/1.
B. L. is supported by a CAPES/COT grant No. 11469/13-17.
A. R. B. acknowledges support from the Department of Physics
and the TCM group of the University of Cambridge as well as
the Cambridge Philosophical Society.

\appendix
\section{An asymptotically AdS Brans-Dicke Black Hole}
\label{app1}

As we have previously discussed in section \ref{sec:conf_transf}, the action \eqref{action} can be brought to the Einstein frame via a conformal transformation. In this frame, the BD action maps to an EMD model. Solutions for this action have been widely studied for different choices of potential \cite{gao2005, sheykhi2010, charmousis2010, kiritsis2015}. For completeness, we give here a particular explicit black brane solution with AdS assymptotics. In the language of ref. \cite{kiritsis2015} this corresponds to a $\delta = \gamma$ solution.
\begin{align}
\label{bdsolution}
\dd \bar{s}^2 &= -f(r)\dd t^2 +\frac{\dd r^2}{f(r)}+r^2 R(r) \delta_{ij}\dd x^i \dd x^j\\
f(r) &= \frac{2\Lambda(\alpha^2+1)^2 b^{2\gamma}}{(d-1)(\alpha^2-d)}r^{2(1-\gamma)}-\frac{m}{r^{(d-1)(1-\gamma)-1}}+\frac{2q^2(\alpha^2+1)^2 b^{-2(d-2)\gamma}}{(d-1)(\alpha^2+d-2)}r^{2(d-2)(\gamma-1)}\\
R(r) &= \left(\frac{b}{r}\right)^{2\gamma} \\
\bar{\phi}(r)&=\frac{(d-1)\alpha}{2(1+\alpha^2)}\log{\frac{b}{r}}\\
\bar{V}(\bar{\phi})&=2\Lambda e^{\frac{4\alpha\bar{\phi}}{d-1}} = \left(\frac{b}{r}\right)^{2\gamma}\\
\bar{a}_t' &= -\frac{q Y}{r^{d-1}}\left(\frac{b}{r}\right)^{-(d-3)\gamma}
\end{align}
\noindent where we defined $\gamma = \alpha^2/(1+\alpha^2)$, with $\alpha$ as in eq. \eqref{eq:coup_conf_transf}. This solution has four free parameters ($\gamma$, b, q, m).  $\gamma$ is a function of the Brans-Dicke parameter $\xi$
\begin{align*}
\gamma = \frac{\alpha^2}{1+\alpha^2} = \frac{1}{1+\alpha^{-2}} = \frac{(d-3)^2}{(d-1)^2+8+4(d-1)\xi}
\end{align*}
The black brane horizon radius is found by imposing $f(r_0)=0$. This allow us to solve for one of the parameters as functions of $r_0$ and the others, e.g. $m=m(r_0,q,b,\gamma)$. The parameter $b$ sets the scale of the dilaton $\bar{\phi}$ and can be set to unit by a coordinate rescaling. 

An interesting case is given by $d=3$, where $\gamma = \alpha = 0$ and the scalar field is constant. In this case the solution above reduces to the well known AdS Reissner-Nordstrom solution. Note that $\gamma=\alpha=0$ also for $\xi \to \infty$, which is the well known Einstein limit of Brans-Dicke theory. In this limit, the solution also reduces to $d+1$ dimensional Reissner-Nordstrom. q is the charge density of the background and $m$ is related to the energy density (ADM mass) as we will discuss below.

To construct an asymptotically AdS black hole solution for our Brans-Dicke theory \eqref{action} we take the inverse conformal mapping on the solution above. The Brans-Dicke black hole is then given by
\begin{align*}
\dd s^2 &= -A(r)\dd t^2 + B(r)\dd r^2 + r^2 c(r) \delta_{ij}\dd x^i \dd x^j \\
A(r) &= \phi^{-\frac{2}{d-1}}f(r)\, \quad 
B(r) = \frac{\phi^{-\frac{2}{d-1}}}{f(r)}\ ,\quad  
c(r) = \phi^{-\frac{2}{d-1}} R(r) = \left(\frac{b}{r}\right)^{\frac{2(d-5)}{d-3}\gamma}\\
V(\phi)&= 2\Lambda\phi^2 \\
a_t' &= -\frac{q Y}{r^{d-1}}\left(\frac{b}{r}\right)^{-(d-3)\gamma}\\
\phi(r) &= \left(\frac{b}{r}\right)^{\frac{2(d-1)\gamma}{d-3}}
\end{align*}
Note that in particular this transformation preserves the position of the horizon $r_0$. The temperature can then be computed
\begin{align*}
4\pi T = |A(r_0)| = \frac{(d-\alpha^2) m}{\alpha^2+1}r_0^{(d-1)(\gamma-1)} - \frac{4q^2(\alpha^2+1)b^{-2(d-2)\gamma}}{\alpha^2+d-2}r_0^{(2d-3)(\gamma-1)-\gamma}
\end{align*}
\noindent which again reduces to the RN temperature for $\gamma=\alpha=0$. The free energy density can be computed by the properly renormalised euclidean action. We refer the curious reader to \cite{SHEYKHI2009b} for the details of the calculation and just quote the answer here,
\begin{align*}
f &= \beta\left(\frac{(d-1)b^{(d-1)\gamma}m}{\alpha^2+1}\right)-\frac{b^{(d-1)\gamma}r_0^{(d-1)(1-\gamma)}}{4\pi} - \beta\frac{q^2}{2((d-3)(1-\gamma)+1)}\frac{1}{r_0^{1+(d-3)(1-\gamma)}}\\
&= \beta \epsilon-s-\beta \mu q
\end{align*}
\noindent for $\beta = T^{-1}$ and we defined
\begin{align*}
\epsilon &= \frac{(d-1)b^{(d-1)\gamma}m}{\alpha^2+1}\ , \quad s =
\frac{b^{(d-1)\gamma}r_0^{(d-1)(1-\gamma)}}{4\pi}\ , \quad \mu =
\frac{q^2}{2((d-3)(1-\gamma)+1)}\frac{1}{r_0^{1+(d-3)(1-\gamma)}}\ .\\
\end{align*}
The energy density (ADM mass), entropy density and chemical potential. In particular note that the above satisfy the first law $\dd \epsilon = T\dd s+\mu \dd \rho $ for $4\pi \rho = \int_{\mathcal{M}} \star F = q$ the charge density.

\section{Gravitational background with $Z\neq 1$, $Y= 1$ and $V\neq 0$ in
four bulk dimensions}\label{appZYV}
\begin{figure}[H]
	\center 
	\includegraphics[scale=0.7,clip]{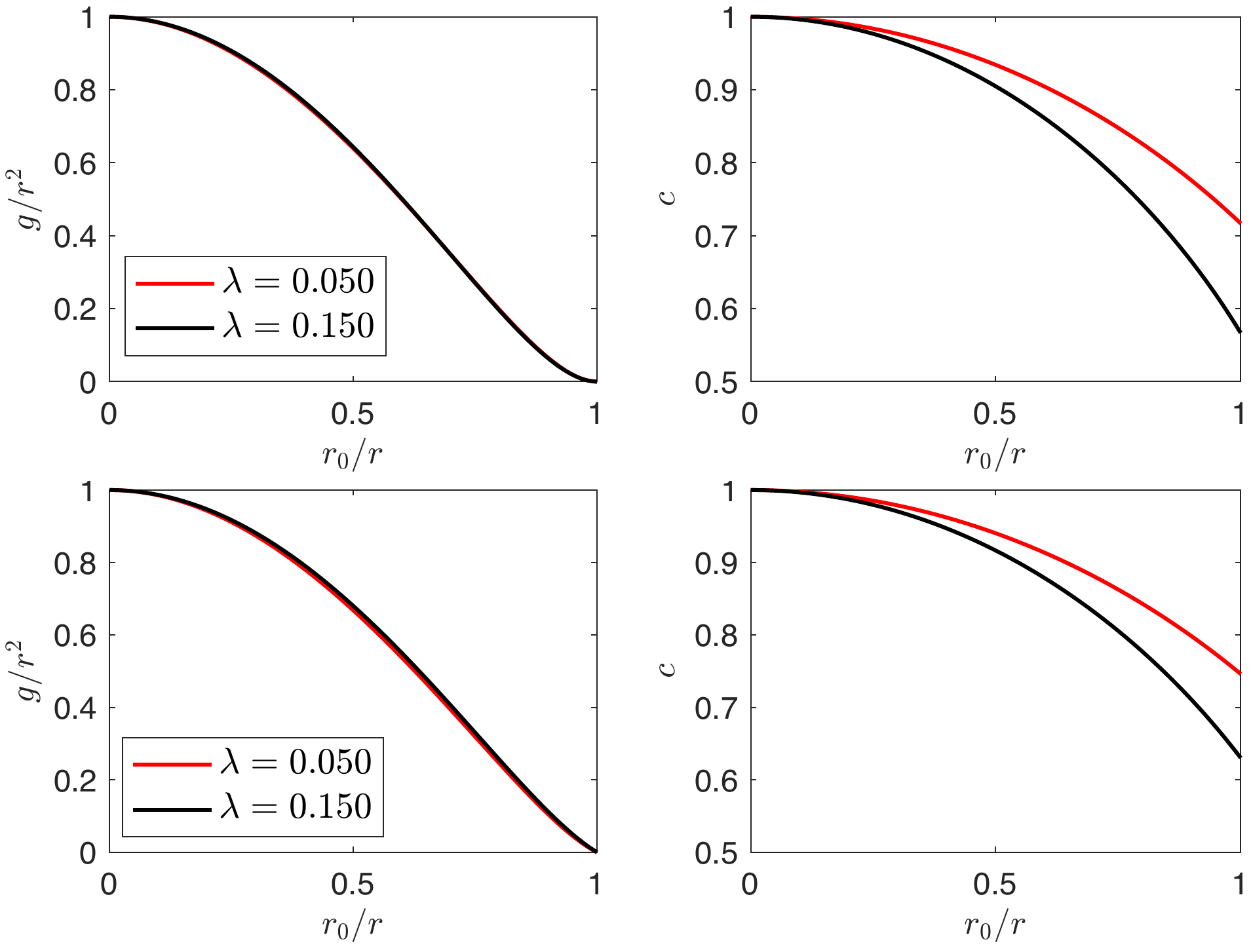}
	\vspace{-0.4cm}
\caption{Metric functions $g$ (blackening factor) and $c$, eq.
(\ref{eq:metric}) for two different temperatures; top row: $T=10^{-4}$ and
bottom row: $T=0.08$.
We fix the charge density $\rho=1$, $\kappa=1$ and
$\lambda_{eff}=\lambda\alpha^2=0.15$.
Each line corresponds to $\alpha^2={\lambda_{eff}/\lambda}$ for the
corresponding $\lambda$, which is given in the legends. The legends also refer
to the right-hand side figures. Moreover, for a fixed temperature and
$\lambda_{eff}$, the dc conductivity is different for the two choices of
$\lambda$ and $\alpha$. We conclude that, in the presence of $V$ in the action,
$\lambda$ and $\alpha$ are two independent parameters associated with the
translational symmetry breaking.}\label{fig:ZYV_background_lameff}
\end{figure}

In Fig.
\ref{fig:ZYV_background_lameff} we show the metric functions, defied in eq.
(\ref{eq:metric}), at nonzero temperature for the model defined in eq.
(\ref{eq:action}) and couplings given in eq. (\ref{eq:couplingsD5}). 

The
lines correspond to a fixed $\lambda_{eff}\equiv \lambda \alpha^2$ but different $\lambda$ and $\alpha$, defined in eq. (\ref{eq:couplings}). Most notably, the function $c$ at the
horizon shows a difference of about $20\%$, while the blackening factor is
very similar throughout the bulk. Fig. \ref{fig:ZYV_background_lameff} shows
that the model defined in eqs. (\ref{eq:action}) and (\ref{eq:couplings})
contains three independent parameters, $\kappa$, $\alpha$ and $\lambda$.

\bibliographystyle{style_ARB}
\bibliography{library}

\end{document}